\documentclass[12pt]{article}
\pdfoutput=1
\usepackage{graphicx}
\usepackage{epstopdf} 
\usepackage{amsmath}
\usepackage{amsfonts}
\usepackage{amssymb}
\usepackage{color}
\usepackage{mathrsfs}
\usepackage[utf8]{inputenc}
\usepackage{subcaption}
\usepackage{placeins}
\usepackage{soul,xcolor}

\usepackage[font=footnotesize]{caption}

\pdfoutput=1

\usepackage[english]{babel}

\usepackage[colorlinks,bookmarks]{hyperref}
\definecolor{linkblue}{rgb}{0,0,0.8}
\definecolor{linkgreen}{rgb}{0,0.5,0}

\hypersetup{pdfpagemode=UseNone, pdfstartview=FitH, linkcolor=linkblue, %
	citecolor=linkgreen, urlcolor=linkblue}

\numberwithin{equation}{section}

\newcommand{\hinvMpc}{\,h\, {\rm Mpc}^{-1}\,}

\newcommand{\bea}{\begin{eqnarray}}
\newcommand{\eea}{\end{eqnarray}}
\newcommand{\be}{\begin{equation}}
\newcommand{\ee}{\end{equation}}
\newcommand{\fr}[2]{\frac{ #1}{#2}}

\newcommand{\kv}{\vec{k}}
\newcommand{\qv}{\vec{q}}

\newcommand{\unitsk}{ h \ { \rm Mpc^{-1}}}

\newcommand{\barr}{\begin{array}}
	\newcommand{\earr}{\end{array}}

\newcommand{\sfrac}[2]{{\textstyle\frac{#1}{#2}}}

\newcommand{\ta}{\tilde{a}}

\newcommand{\mG}{\mathcal{G}}

\newcommand{\mU}{\mathcal{U}}
\newcommand{\mV}{\mathcal{V}}

\newcommand{\vk}{\vec{k}}
\newcommand{\vq}{\vec{q}}

\newcommand{\vx}{\vec{x}}

\usepackage[colorlinks,bookmarks]{hyperref}
\definecolor{linkblue}{rgb}{0,0,0.8}
\definecolor{linkgreen}{rgb}{0,0.5,0}

\hypersetup{pdfpagemode=UseNone, pdfstartview=FitH, linkcolor=linkblue, %
	citecolor=linkgreen, urlcolor=linkblue}

\def\beq{\begin{equation}}
\def\eeq{\end{equation}}
\def\be{\begin{equation}}
\def\ee{\end{equation}}
\def\bea{\begin{eqnarray}}
\def\eea{\end{eqnarray}}
\def\d{{\partial}}

\def\km{{k_{\rm M}}}

\def\xfl{{\vec x_{\rm fl}}}

\newcommand{\delhr}{\delta_{h,r}}
\newcommand{\delh}{\delta_{h}}

\newcommand{\q}{\vec{q}}

\newcommand{\dzdsq}{\fr{\partial_z}{\partial^2}}

\newcommand{\xv}{\vec{x}}

\newcommand{\sym}{{\rm sym}}

\newcommand{\Comment}[1]{{}}

\begin{document}

	\setstcolor{red}

	\setcounter{page}{1} \baselineskip=15.5pt \thispagestyle{empty}
	
	\begin{flushright}
	\end{flushright}
	
	\begin{center}

		{\Large \bf Biased Tracers in Redshift Space in the\\[0.3cm] EFTofLSS with exact time dependence \\[0.7cm]}
		{\large Yaniv Donath$^{1,2,3}$, Leonardo Senatore$^{2,3}$}
		\\[0.7cm]
		
		{\normalsize { \sl $^{1}$ Institute for Theoretical Physics, ETH Zurich, 8093 Zürich, Switzerland
		}}\\
		\vspace{.3cm}
		
		{\normalsize { \sl $^{2}$ Stanford Institute for Theoretical Physics,\\ Stanford University, Stanford, CA 94306}}\\
		\vspace{.3cm}
		
		{\normalsize { \sl $^{3}$ Kavli Institute for Particle Astrophysics and Cosmology and Dept. of Particle Physics and Astrophysics, SLAC, Menlo Park, CA 94025}}\\
		\vspace{.3cm}

	\end{center}
	
	\vspace{.8cm}

	%
	%
	%
	
	\hrule \vspace{0.3cm}
	{\small \noindent \textbf{Abstract} \\[0.3cm]
		We study the effect of the Einstein -- de Sitter (EdS) approximation on the one-loop power spectrum of galaxies in redshift space in the Effective Field Theory of Large-Scale Structure. The dark matter density perturbations and velocity divergence are treated with exact time dependence. Splitting the density perturbation into its different temporal evolutions naturally gives rise to an irreducible basis of biases. While, as in the EdS approximation, at each time this basis spans a seven-dimensional space, this space is a slightly different one, and the difference is captured by a single calculable time- and $\vec k$-dependent function. 
		We then compute the redshift-space galaxy one-loop power spectrum with the EdS approximation~($P^{\text{EdS-approx}}$) and without~($P^{\text{Exact}}$). For the monopole we find $P_{\text{0}}^{\text{Exact}}/P_{\text{0}}^{\text{EdS-approx}}\sim 1.003$ and for the quadrupole $P_{\text{2}}^{\text{Exact}}/P_{\text{2}}^{\text{EdS-approx}}\sim 1.007$ at $z=0.57$, and sharply increasing at lower redshifts. Finally, we show that a substantial fraction of the effect remains even after allowing the bias coefficients to shift within a physically allowed range. This suggests that the EdS approximation can only fit the data to a level of precision that is roughly comparable to the precision of the next generation of cosmological surveys. Furthermore, we find that implementing the exact time dependence formalism is not demanding and is easily applicable to data. Both of these points motivate a direct study of this effect on the cosmological parameters.
		\noindent 
		\vspace{0.3cm}
		\hrule
		
		\def\thefootnote{\arabic{footnote}}
		\setcounter{footnote}{0}

		\newpage
		%
		%
		%
		%
		\tableofcontents
		 \vspace{1in}
		\section{Introduction}
		The recent analysis of SDSS/BOSS data~\cite{DAmico:2019fhj,Ivanov:2019pdj,Colas:2019ret,Ivanov:2019hqk,Philcox:2020vvt,DAmico:2020kxu} using the Effective Field-Theory of Large-Scale Structure (EFTofLSS)~\cite{Baumann:2010tm,Carrasco:2012cv,Porto:2013qua,Senatore:2014via} has provided the first CMB-independent low-redshift measurement of $H_0$ in agreement with Planck~\cite{Aghanim:2018eyx}. In addition, it allows us to measure all of the cosmological parameters only using a Big Bang Nucleosynthesis prior on
		the fractional energy density of baryons $\Omega_b h^2$. With the increasing precision of current and upcoming large-scale structure surveys, we will probably be able to tighten the error bars to sub-percent levels. In turn, this means that the theory has to hold to a similar level of precision.
		
		There have been several developments in order to tackle the challenge of reaching sub percent precision. One of them are numerical simulations, which try to model the formation of galaxies, on top of exact simulations of the underlying dark matter fields. While this approach has brought about some very important results, it is unclear how scalable it is and thus there have been limits to the amount of information we have been able to extract with this method from large-scale structure surveys. 
		
		Over the past few years, the exact, analytical side has made a lot of progress in the form of cosmological perturbation theory. Specifically, the EFTofLSS has been in remarkable agreement with both data and simulations. The EFTofLSS is a perturbative framework to calculate large-scale structure correlation functions in the mildly non-linear regime~\cite{Baumann:2010tm,Carrasco:2012cv,Porto:2013qua,Senatore:2014via,Carrasco:2013sva,Carrasco:2013mua,Pajer:2013jj,Carroll:2013oxa,Mercolli:2013bsa,Angulo:2014tfa,Baldauf:2014qfa,Senatore:2014eva,Senatore:2014vja,Lewandowski:2014rca,Mirbabayi:2014zca,Foreman:2015uva,Angulo:2015eqa,McQuinn:2015tva,Assassi:2015jqa,Baldauf:2015tla,Baldauf:2015xfa,Foreman:2015lca,Baldauf:2015aha,Baldauf:2015zga,Bertolini:2015fya,Bertolini:2016bmt,Assassi:2015fma,Lewandowski:2015ziq,Cataneo:2016suz,Bertolini:2016hxg,Fujita:2016dne,Lewandowski:2016yce,Lewandowski:2017kes,Senatore:2017hyk,deBelsunce:2018xtd,Perko:2016puo}. It captures the effect that UV-physics has on the long-wavelength observables, by including additional terms in the equations of motion for those wavelengths. Up to now, the EFTofLSS predicts the correlation functions of dark matter~\cite{Carrasco:2012cv,Senatore:2014via,Carrasco:2013sva,Carrasco:2013mua,Angulo:2014tfa,Baldauf:2014qfa,Senatore:2014vja,Foreman:2015lca,Baldauf:2015aha,Bertolini:2015fya,Bertolini:2016bmt} and biased tracers~\cite{Senatore:2014vja,Angulo:2015eqa,Fujita:2016dne}, including the presence of dark energy~\cite{Lewandowski:2016yce,Lewandowski:2017kes} and massive neutrinos~\cite{Senatore:2017hyk,deBelsunce:2018xtd}. 
		
		In principle, the EFTofLSS has the potential to reach extremely high accuracy in the mildly non-linear regime, {\it i.e.} $0.1h\, {\rm{Mpc}}^{-1} \lesssim k_{\rm max }\lesssim0.5h\, {\rm{Mpc}}^{-1}$, by going to arbitrary high orders in perturbations. In practice, we of course compute observables only to some finite order, which in our case is up to one loop. At this level it has recently been shown \cite{DAmico:2019fhj} that we can trust the prediction for the power spectrum ({\it i.e.} the theoretical error is negligible) up to $k\sim0.2h\, \rm{Mpc}^{-1}$.
		
		In order to mathematically facilitate perturbative calculations often the Einstein -- de Sitter (EdS) approximation is used~\cite{Senatore:2014eva,Mirbabayi:2014zca,Perko:2016puo}. It is inspired by the fact that in an EdS cosmology (where the fractional matter density is $\Omega_m$ = 1, and there is no dark energy $\Omega_D$ = 0), the time dependence of density perturbations goes as $ D^n(a)$, where $D$ is the growth factor. It is thus tempting to use this identity in a more complicated cosmology, such as $\Lambda$CDM or $w$CDM. As has been shown in~\cite{Takahashi:2008yk,Pietroni:2008jx,Carrasco:2012cv,Lewandowski:2016yce,Lewandowski:2017kes, Fasiello:2016qpn}, using the EdS approximation in our universe is accurate to percent level precision on the full power spectrum in real space. Yet, current and upcoming low-redshift surveys may come increasingly close to this threshold, where the EdS approximation might no longer be precise enough, especially in redshift space. It is, therefore, necessary to extend our theory to an exact time dependence, in order to at least check the validity of the approximation in redshift space. In practice, this means that the different momentum kernels will evolve separately in time and not with a common factor of $D^n(a)$.
		
		In this paper, we extend the theory of biased tracers in redshift space, formulated in~\cite{Perko:2016puo} and applied to data in~\cite{DAmico:2019fhj}, to an exact time dependence. This entails the revision and extension of the EFTofLSS at several steps. We start with the exact time dependence for the dark matter density field as developed in~\cite{Lewandowski:2016yce}, which results in the separate time evolution of the momentum kernels. We then generalize the treatment of biased tracers from~\cite{Senatore:2014eva,Mirbabayi:2014zca} to include the exact time dependence of the dark matter density fields. Similarly to~\cite{Angulo:2015eqa}, we find that the ad hoc treatment of the bias will lead to degeneracies and to a too large number of bias coefficients. In the context of resolving this degeneracy, we introduce an alternative basis to former ones~\cite{Angulo:2015eqa,Mirbabayi:2014zca}, which comes naturally from the momentum kernels appearing in the density perturbations. In a last step, as has been developed in~\cite{Perko:2016puo}, we do the transformation to redshift space. To compute the one-loop halo power spectrum in redshift space we introduce counter-terms to renormalize the biases in real and redshift space. However, the exact time dependence does not change the counter-terms, and the theory does not have to be extended in this part.
		
		The purpose of this paper is to determine the validity of the EdS approximation in the case of the one-loop power spectrum in redshift space for biased tracers. We make the relevant theoretical developments to calculate the one-loop halo power spectrum in redshift space with exact time dependence and use the recently measured parameters from~\cite{DAmico:2019fhj} to estimate the impact of the EdS approximation. Furthermore, we check whether a change in the bias coefficients in the EdS approximation can account for this impact. We will leave the measurement of the bias coefficients in the exact case to future work.
		
		%
		%
		%
		%
		\section{Biased tracers with exact time dependence}\label{sec:biasedexact}
		The halo overdensity depends on the underlying distribution of dark matter, therefore we start with the continuity, Euler and Poisson equation for the dark matter field 
		\bea\label{eq:masterreal1}
		&&aH\delta' + \frac{1}{a} \partial_i ( (1 + \delta)v^i) =0 \\
		\label{fluid2} 
		\label{eq:masterreal2}
		&&aH\partial_i {v'}^i + H \partial_i v^i + \frac{1}{a} \partial_i ( v^j \partial_j v^i ) + \frac{1}{a} \partial^2 \Phi = - \frac{1}{a} \partial_i \left( \frac{1}{\rho} \partial_j \tau^{ij} \right) \\
		\label{eq:masterreal3}
		&& a^{-2} \partial^2 \Phi = \frac{3}{2 } \frac{ \Omega_{m,0} H_0^2 a_0}{a^3} \delta \label{poissonfish}\ ,
		\eea
		where $\Phi$ is the gravitational potential, $\delta$ the density perturbation, $v$ the peculiar velocity field, $\rho$ the background density and $\tau^{ij}$ the effective stress-tensor responsible for the counter-terms discussed in section~\ref{countstoch}. We use the scale factor $a$ as our time variable such that $' = \partial/\partial a$ and $a_0$ is the present-day scale factor which from here on we set to unity. The equations of motion in the EFTofLSS in Fourier space without the counter-terms are~\cite{Bernardeau:2001qr}
		\bea\label{eq:master1}
		&&a\delta'_{\vk}-f_{+}\theta_{\vk}=(2\pi)^{3}f_{+}\iint \frac{d^3q_1}{(2\pi)^{3}}\frac{d^3q_2}{(2\pi)^{3}} \delta_{D}(\vk-\vq_1-\vq_2)\alpha(\vq_1,\vq_2)\theta_{\vq_1}\delta_{\vq_2},\\
		&&a\theta'_{\vk}-f_{+}\theta_{\vk}+\fr{3}{2} \frac{\Omega_{m}}{f_{+}}(\theta_{\vk}-\delta_{\vk})=(2\pi)^{3}f_{+}\iint \frac{d^3q_1}{(2\pi)^{3}}\frac{d^3q_2}{(2\pi)^{3}}\delta_{D}(\vk-\vq_1-\vq_2)\beta(\vq_1,\vq_2)\theta_{\vq_1}\theta_{\vq_2} 
		\label{eq:master2},
		\eea
		where as usual $\alpha(\vq_1,\vq_2) = 1+\fr{\vq_1\cdot\vq_2}{q_1^2}$, $\beta(\vq_1,\vq_2) = \fr{(\vq_1+\vq_2)^2\vq_1\cdot\vq_2}{2q_1^2q_2^2}$ and $\delta_{D}$ is the delta distribution. To linear order we have $\delta = \theta$, where $\theta = -\fr{1}{f_{+}aH}\partial_iv^i$ is the rescaled velocity divergence, $\Omega_{m}$ is the time-dependent fractional matter density and $f_{+} = \fr{aD_{+}'}{D_{+}}$ is the linear growth rate in terms of the growth factor (see Appendix~\ref{appendixa}). We write the dark matter overdensities and velocity divergence in a perturbative expansion of the form
		\be \label{deltaexpansion}
		\delta_{\vk}(a)=\sum^{\infty}_{n=1}\delta_{\vk}^{(n)}(a) \quad \textmd{and} \quad \theta_{\vk}(a) =\sum^{\infty}_{n=1} \theta_{\vk}^{(n)}(a),
		\ee
		which allows us to solve equations (\ref{eq:master1}) and (\ref{eq:master2}) order by order. The full solutions to the dark matter overdensities and velocity divergence also includes the dark matter field counter-terms $\delta_{\vk}^{(ct)}$ and $\theta_{\vk}^{(ct)}$, which we will discuss in section~\ref{countstoch}. The perturbative solutions in (\ref{deltaexpansion}) can generally be written as an integral over time-dependent momentum kernels
		\bea
		\label{eq:kernelform}
		\delta^{(n)}_{\vk}(a)=\int\frac{d^3q_1}{(2\pi)^{3}}...\frac{d^3q_n}{(2\pi)^{3}}(2\pi)^{3}\delta_{D}(\vk-\vq_1-...-\vq_n)K_\delta^{(n)}(\vq_1,...,\vq_n,a)\delta^{(1)}_{\vq_1}(a)...\delta^{(1)}_{\vq_n}(a) \\ \nonumber
		\theta^{(n)}_{\vk}(a)=\int\frac{d^3q_1}{(2\pi)^{3}}...\frac{d^3q_n}{(2\pi)^{3}}(2\pi)^{3}\delta_{D}(\vk-\vq_1-...-\vq_n)K_\theta^{(n)}(\vq_1,...,\vq_n,a)\delta^{(1)}_{\vq_1}(a)...\delta^{(1)}_{\vq_n}(a).
		\eea
		
		In section~\ref{halos} we are going to expand the halo overdensity up to third order in perturbations, using exact time dependence. The halo overdensity at a given order depends on the dark matter fields up to that same order. Therefore, we here give the time-dependent kernels of the dark matter fields, i.e. solutions to (\ref{eq:master1}) and (\ref{eq:master2}) (see for example~\cite{Lewandowski:2016yce}, {setting $C=1$ and using the growth factor of a $w$CDM cosmology}), up to cubic order
		\bea\label{kernels1}
		K_\lambda^{(1)}(\vq_1,a)&=& 1 \\
		\label{kernels2}
		K_\lambda^{(2)}(\vq_1,\vq_2,a)&=&\alpha_s(\vq_1,\vq_2)\mG^{\lambda}_{1}(a)+\beta(\vq_1,\vq_2)\mG^{\lambda}_{2}(a) \\
		\label{kernels3}
		K_\lambda^{(3)}(\vq_1,\vq_2,\vq_3,a)&=&\alpha^{\sigma}(\vq_1,\vq_2,\vq_3)\mU^{\lambda}_{\sigma}(a) +\beta^{\sigma}(\vq_1,\vq_2,\vq_3)\mV^{\lambda}_{\sigma2}(a) + \gamma^{\sigma}(\vq_1,\vq_2,\vq_3)\mV^{\lambda}_{\sigma1}(a),
		\eea
		where repeated $\sigma \in \{1,2\}$ are summed over and $\lambda \in \{\delta,\theta\}$. For simplicity we symmetrized $\alpha$ to $\alpha_s(\vq_1,\vq_2) = \fr{1}{2}(\alpha(\vq_1,\vq_2)+\alpha(\vq_2,\vq_1))$ and the six momentum kernels at third order $\{\alpha_\sigma,\beta_\sigma,\gamma_\sigma\}$ are products of $\alpha_s$ and $\beta$ given in Appendix~\ref{appendixb}. $\{\mG^{\lambda}_1,\mG^{\lambda}_2,\mU^{\lambda}_{\sigma},\mV^{\lambda}_{\sigma\tilde{\sigma}}\}$, where $\tilde{\sigma} \in \{1,2\}$, are time-dependent functions resulting from equations (\ref{eq:master1}) and (\ref{eq:master2}). They are explicitly given in Appendix~\ref{appendixa}.

		\subsection{Perturbative expansions of $\delta_h$ and $\theta_h$ }
		\label{halos}
		We are interested in the bias expansion of the halo density fluctuations, which depends only on the dark matter field and its derivatives allowed by the equivalence principle. Following the notation of~\cite{Senatore:2014eva} the expansion is given in Eulerian space by
		\bea\label{eq:euler_bias_raw}
		&&\delta_h(\vec x,a)\simeq \int^a \fr{da'}{a'}\; \left[ c_{\delta}(a,a')\; :\delta(\xfl,a'): \right.\\ \nonumber
		&&\qquad+ c_{\delta^2}(a,a')\; :\delta(\xfl,a')^2: + c_{s^2}(a,a')\; :s^2(\xfl,a'):\\\nonumber
		&&\qquad+ c_{\delta^3}(a,a')\; :\delta(\xfl,a')^3 : + c_{\delta s^2}(a,a')\; : \delta(\xfl,a') s^2(\xfl,a'):+ c_{\psi}(a,a')\; :\psi(\xfl,a'):\\ \nonumber
		&&\qquad\qquad+ c_{\delta st}(a,a') \; :\delta(\xfl,a') st(\xfl,a'):+ c_{ s^3}(a,a')\; :s^3(\xfl,a'):\\\nonumber
		&&\qquad+ c_{\epsilon}(a,a')\;\epsilon(\xfl,a')\\ \nonumber
		&&\qquad+ c_{\epsilon\delta}(a,a') \;:\epsilon(\xfl,a')\delta(\xfl,a'):+ c_{\epsilon s}(a,a') \;:\epsilon s(\xfl,a'):+ c_{\epsilon t }(a,a') \;:\epsilon t(\xfl,a'):\\\ \nonumber
		&&\qquad+ c_{\epsilon^2\delta}(a,a') \;:\epsilon(\xfl,a')^2\delta(\xfl,a'):
		+ c_{\epsilon\delta^2}(a,a') \;:\epsilon(\xfl,a')\delta(\xfl,a')^2:+ c_{\epsilon s^2}(a,a') \;:\epsilon(\xfl,a')s^2(\xfl,a'): \\ \nonumber
		&&\qquad\qquad+ c_{\epsilon s \delta}(a,a') \;:\epsilon s(\xfl,a')\delta(\xfl,a'):+ c_{\epsilon t \delta}(a,a') \;:\epsilon t(\xfl,a')\delta(\xfl,a'):\\ \nonumber
		&&\left.\qquad+ c_{\d^2\delta}(a,a')\; \;\frac{\d^2_{x_{\rm fl}}}{\km^2}\delta(\xfl,a')+\dots\ \right],
		\eea
		where $\km$ is the comoving wavenumber that encloses the mass of the galaxy and $\epsilon(\vx,a)$ is the stochastic field that accounts for the difference between a given realization and the average of the dark matter field. Both of these terms are discussed in section~\ref{countstoch}.\\
		The above expansion is normal ordered, i.e. $ :{\cal O}:=\cal O-\langle\cal O\rangle$. Furthermore, we recursively define $\xfl$
		\bea\label{xfl}
		\xfl(\vec x, a,a') = \vec x - \int_{a'}^{a} \fr{da''}{a''^2H(a'')} \vec{v}(a'',\vec x_{\rm fl}(\vec x, a,a'')).
		\eea
		We find it useful to define quantities that only start at second order~\cite{McDonald:2009dh}, such as 
		\be\label{eta}
		\eta(\vec x,t)=\theta(\vec x,t)- \delta(\vec x,t)\ .
		\ee
		The tidal tensor $s_{ij}$ and $t_{ij}$ are defined as
		\beq 
		s_{ij}(\vec{x},a)=\mathcal{D}_{ij}\delta(\vec{x},a) \quad \textmd{and} \quad t_{ij}(\vec{x},a)=\mathcal{D}_{ij}\eta(\vec{x},a),
		\eeq
		where $\mathcal{D}_{ij} = \fr{\partial_i \partial_j}{\partial^2}-\fr{1}{3}\delta_{ij}$. The non-vanishing contractions of these operators, appearing in equation (\ref{eq:euler_bias_raw}) are defined as
		\bea\label{eq:operators}
		&&s^2(\xfl,a)= s_{ij}(\xfl,a) s^{ij}(\xfl,a)\ , \quad s^3(\xfl,a)=s_{ij}(\xfl,a) s^{il}(\xfl,a)s_l{}^{j}(\xfl,a)\ ,\\ \nonumber
		&& st(\xfl,a)=s_{ij}(\xfl,a) t^{ij}(\xfl,a)\ , \quad \epsilon s(\xfl,a)=\epsilon_{ij}(\xfl,a)s^{ij}(\xfl,a), \quad \epsilon t(\xfl,a)=\epsilon_{ij}(\xfl,a)t^{ij}(\xfl,a)\ ,
		\eea
		where indices are raised with $\delta^{ij}$. Similarly to the construction of $\psi$ in~\cite{McDonald:2009dh}, we want $\psi$ only to start at cubic order. We notice that
		\bea\label{eta2}
		\eta(\vec{x},a)^{(2)} = \left(\mG^{\delta}_{1}(a)-\mG^{\theta}_{1}(a)\right)\left(s^2{}^{(2)}(\vec{x},a)-\frac{2}{3}\delta^2{}^{(2)}(\vec{x},a)\right),
		\eea which follows from equations (\ref{kernels2}), (\ref{eta}) and the fact that $\mG^{\delta}_1+\mG^{\delta}_2 = \mG^{\theta}_1+\mG^{\theta}_2$, which is shown in Appendix~\ref{appendixa}. In the EdS approximation, equation (\ref{eta2}), reduces to $\eta^{(2)}=\frac{2}{7}s^2{}^{(2)}-\frac{4}{21}\delta^2{}^{(2)}$~\cite{McDonald:2009dh}, because $\mG^{\delta}_1(a)\stackrel{EdS}{=}\fr{5}{7}$ and $\mG^{\theta}_1(a)\stackrel{EdS}{=}\fr{3}{7}$ in said approximation. Following the construction above, $\psi$ is given by
		\bea 
		\psi(\vec{x},a) = \theta(\vec{x},a)-\delta(\vec{x},a)-\left(\mG^{\delta}_{1}(a)-\mG^{\theta}_{1}(a)\right)\left(s^2(\vec{x},a)-\frac{2}{3}\delta^2(\vec{x},a)\right),
		\eea
		and will only start at cubic order. 
		
		As has been pointed out in~\cite{Angulo:2015eqa}, the operators in equation (\ref{eq:euler_bias_raw}) are degenerate at a given, low, order. We, therefore, face two challenges. First, we cannot perform the time integrals symbolically as has been done in~\cite{Senatore:2014eva}, without expanding $\delta$ into its different temporal evolutions. Secondly, we have to find an irreducible basis for the biases. 
		
		We start with equation (\ref{eq:euler_bias_raw}) in Fourier space. Note that the expressions are evaluated at $\xfl$, and we, therefore, Taylor expand $\delta( \xfl,a)$ up to cubic order, which is given by
		\bea\label{eq:xfl_expansion}
		&&\delta(\xfl(a,a'),a')=\delta(\vec x,a')-\d_i\delta(x,a')\int_{a'}^a 	\fr{da''}{a''^2H(a'')} \; v^i(\vec x,a'')\\ \nonumber
		&&\quad \quad\quad\quad\quad\quad+\frac{1}{2}\d_i\d_j \delta(x,a')\int_{a'}^a 	\fr{da''}{a''^2H(a'')}\; v^i(\vec x,a'')\int_{a'}^a 	\fr{da'''}{a'''^2H(a''')}\;v^j(\vec x,a''')\\ \nonumber
		&&\quad \quad\quad \quad\quad \quad+\d_i\delta(x,a')\int_{a'}^a 	\fr{da''}{a''^2H(a'')}\; \d_j v^i(\vec x,a'') \int^a_{a''} 	\fr{da'''}{a'''^2H(a''')}\;v^j(\vec x,a''')+\ldots\ .
		\eea 
		The halo overdensities in Fourier space therefore read
		\bea\label{eq:euler_bias_new}
		&&\delta_h(\vk,a)\quad=\quad \\ \nonumber
		&&\quad = c_{\delta,1}(a) \delta^{(1)}(\vk,a)+ \int^a \frac{da'}{a'}c_{\delta}(a,a') \delta^{(2)}(\vk,a')+ \int^a \frac{da'}{a'}c_{\delta}(a,a') \delta^{(3)}(\vk,a')+ \int^a \frac{da'}{a'}c_{\delta}(a,a') \delta^{(3)}_{c_t}(\vk,a')\\ \nonumber
		&&\qquad+ c_{\delta,12}(a) [\d_i \delta^{(1)}\; \frac{\d^i}{\d^2}\theta^{(1)}]_{\vk}(a)+ \int^a \frac{da'}{a'}\;c_{\delta}(a,a') \left[1-\frac{D_{+}(a')}{D_{+}(a)}\right][\d_i \delta^{(2)}(a')\;\frac{\d^i}{\d^2}\theta^{(1)}(a)]_{\vk}\\ \nonumber
		&&\qquad+ \int^a \frac{da'}{a'}c_{\delta}(a,a')\frac{D_{+}(a')}{D_{+}(a)}\int_{a'}^a da''\frac{D'_{+}(a'')}{D_{+}(a'')}\;[\d_i \delta^{(1)}(a)\;\frac{\d^i}{\d^2}\theta^{(2)}(a'')]_{\vk} \\ \nonumber
		&&\qquad+c_{\delta,123}(a)
		\left[ [\d_i \delta^{(1)}\; \frac{\d_j\d^i}{\d^2}\theta^{(1)}\; \frac{\d^j}{\d^2}\theta^{(1)}]_{\vk}(a)+[\d_i\d_j \delta^{(1)}\;\frac{\d^i}{\d^2}\theta^{(1)}\frac{\d^j}{\d^2}\theta^{(1)}]_{\vk}(a)\right]+ \\ \nonumber
		&&\quad+c_{\delta^2,1}(a) \; [\delta^2]^{(2)}_{\vk}(a)+2\int^a \frac{da'}{a'}c_{\delta^2}(a,a')\;\frac{D_{+}(a')}{D_{+}(a)} \;\; [\delta^{(1)}(a)\delta^{(2)}(a')]_{\vk}\\ \nonumber
		&& \qquad+2 c_{\delta^2,12}(a)[\delta^{(1)}\d_i\delta^{(1)}\frac{\d^i}{\d^2}\theta^{(1)}]_{\vk}(a) \\ \nonumber
		&& \quad
		+c_{s^2,1}(a) \; [s^2]_{\vk}^{(2)}(a)+2\int^a \frac{da'}{a'}\;c_{s^2}(a,a')\frac{D_{+}(a')}{D_{+}(a)}\;\left[[\frac{\d_i\d_j}{\d^2}\delta^{(2)}(a')\frac{\d^i\d^j}{\d^2}\delta^{(1)}(a)]_{\vk}-\frac{1}{3}[\delta^{(2)}(a')\delta^{(1)}(a)]_{\vk}\right] \\ \nonumber
		&& \qquad+2 c_{s^2,12}(a)[s_{lm}^{(1)}\d_i (s^{lm})^{(1)}\frac{\d^i}{\d^2}\theta^{(1)}]_{\vk}(a)\\ \nonumber
		&&\quad +\int^a \frac{da'}{a'}c_{st}(a,a')\frac{D_{+}(a')}{D_{+}(a)}\left([\frac{\d_i\d_j}{\d^2}\delta^{(1)}(a)\frac{\d^i\d^j}{\d^2}\theta^{(2)}(a')]_{\vk}-\frac{1}{3}[\delta^{(1)}(a)\theta^{(2)}(a')]_{\vk}\right) \\ \nonumber
		&&\qquad +\int^a \frac{da'}{a'}c_{st}(a,a')\frac{D_{+}(a')}{D_{+}(a)}\left([\frac{\d_i\d_j}{\d^2}\delta^{(2)}(a')\frac{\d^i\d^j}{\d^2}\theta^{(1)}(a)]_{\vk}-\frac{1}{3}[\delta^{(2)}(a')\theta^{(1)}(a)]_{\vk}\right) \\ \nonumber
		&&\qquad -2\int^a \frac{da'}{a'}c_{st}(a,a')\frac{D_{+}(a')}{D_{+}(a)}\left([\frac{\d_i\d_j}{\d^2}\delta^{(1)}(a)\frac{\d^i\d^j}{\d^2}\delta^{(2)}(a')]_{\vk}-\frac{1}{3}[\delta^{(1)}(a)\delta^{(2)}(a')]_{\vk}\right) \; \\ \nonumber 
		&&\quad +\int^a\frac{da'}{a'}c_{\psi}(a,a')[\psi^{(3)}]_{\vk}(a') \; \\ \nonumber
		&&\quad+ c_{\delta^3}(a)[\delta^3]^{(3)}_{\vk}(a)+c_{\delta\,s^2}(a)[\delta s^2]^{(3)}_{\vk}(a)
		+c_{s^3}(a)[s^3]^{(3)}_{\vk}(a)+c_{\delta\,\epsilon^2}(a)[\delta \epsilon^2]^{(3)}_{\vk}(a)\\ \nonumber
		&&\quad+ c_{\epsilon,1}(a) \;[\epsilon^{(1)}]_{\vk}+ c_{\epsilon,2} \;[\epsilon^{(2)}]_{\vk}+\ldots\ .
		\eea
		Let us explain the structure of the above expansion. In the first line, we have the density perturbation up to third order, including the speed of sound counter-term. Lines two to four are due to the flow terms that stem from equation (\ref{eq:xfl_expansion}). Similarly, the rest of the terms are followed by possible flow terms, derived in Appendix~\ref{appendixc}. 
		
		For expressions that are convolutions of $\delta^{(1)}$ we are able to do the time integral symbolically, i.e. we absorb them into coefficients, such as
		\be
		\label{eq:coefs}
		c_{\delta,1}(a)= \int^a \frac{da'}{a'}c_{\delta}(a,a')\frac{D_{+}(a')}{D_{+}(a)} , \qquad c_{\delta^2,1}(a)=\int^a \frac{da'}{a'}c_{\delta^2}(a,a')\frac{D_{+}(a')^2}{D_{+}(a)^2} ,\quad \ldots\ .
		\ee
		All other products that consist of perturbations of order two or higher, must be expanded into their various temporal evolutions. However, we can recognize that all the mode-dependent terms in equation (\ref{eq:euler_bias_new}) share structure. For example, we can write the flow term in the second line, in terms of the kernel $\alpha$ that appears at the second order of the density perturbation
		\be
		[\d_i \delta^{(1)}\; \frac{\d^i}{\d^2}\theta^{(1)}]_{\vk}(a) = \int\frac{d^3q_1}{(2\pi)^{3}}\frac{d^3q_2}{(2\pi)^{3}}(2\pi)^{3}\delta_{D}(\vk-\vq_1-\vq_2)(\alpha(\vq_1,\vq_2)-1) \delta^{(1)}_{\vq_1}(a)\delta^{(1)}_{\vq_2}(a).
		\ee
		More generally, we can write all terms in (\ref{eq:euler_bias_new}) (neglecting the stochastic and the counter-terms for now) as integrals over the nine momentum functions $\{1,\alpha,\beta,\alpha_1,\alpha_2,\beta_1,\beta_2,\gamma_1,\gamma_2\}$ that appear in equations (\ref{kernels1})-(\ref{kernels3}) (See (\ref{eq:euler_bias_kints}) in Appendix~\ref{appendixb}). Next, we collect the temporal coefficients into thirteen parameters and obtain the halo density kernels
		\bea\label{eq:euler_bias_k}
		K_{\delta_h}^{(1)}(\vq_1,a)&=& c_{\delta,1}(a) \\ \nonumber 
		K_{\delta_h}^{(2)}(\vq_1,\vq_2,a)&=&c_{\mathbb{I},(2)}(a)+c_{\alpha,(2)}(a)\alpha(\vq_1,\vq_2)+c_{\beta,(2)}(a)\beta(\vq_1,\vq_2) \\ \nonumber
		K_{\delta_h}^{(3)}(\vq_1,\vq_2,\vq_3,a)&=&c_{\alpha_\sigma,(3)}(a)\alpha^\sigma(\vq_1,\vq_2,\vq_3) +c_{\beta_\sigma,(3)}(a)\beta^\sigma(\vq_1,\vq_2,\vq_3)+c_{\gamma_\sigma,(3)}(a)\gamma^\sigma(\vq_1,\vq_2,\vq_3)\\ \nonumber
		&&+c_{\alpha,(3)}(a)\alpha(\vq_1,\vq_2)+c_{\beta,(3)}(a)\beta(\vq_1,\vq_2) +c_{\mathbb{I},(3)}(a),
		\eea
		where $\sigma \in \{1,2\}$, repeated indices are summed over and the halo kernels are similarly to (\ref{eq:kernelform}) defined at each order
		\bea
		\label{eq:kernelformh}
		\delta^{(n)}_{h}(\vk,a)=\int\frac{d^3q_1}{(2\pi)^{3}}...\frac{d^3q_n}{(2\pi)^{3}}(2\pi)^{3}\delta_{D}(\vk-\vq_1-...-\vq_n)K_{\delta_h}^{(n)}(\vq_1,...,\vq_n,a)\delta^{(1)}_{\vq_1}(a)...\delta^{(1)}_{\vq_n}(a) \\ \nonumber
		\theta^{(n)}_{h}(\vk,a)=\int\frac{d^3q_1}{(2\pi)^{3}}...\frac{d^3q_n}{(2\pi)^{3}}(2\pi)^{3}\delta_{D}(\vk-\vq_1-...-\vq_n)K_{\theta_h}^{(n)}(\vq_1,...,\vq_n,a)\delta^{(1)}_{\vq_1}(a)...\delta^{(1)}_{\vq_n}(a).
		\eea
		The coefficients $\{c_{\delta,1},c_{\mathbb{I},(2)},c_{\alpha,(2)},c_{\beta,(2)},c_{\mathbb{I},(3)},c_{\alpha,(3)},c_{\beta,(3)},c_{\alpha_1,(3)},c_{\alpha_2,(3)},c_{\beta_1,(3)},c_{\beta_2,(3)},c_{\gamma_1,(3)},c_{\gamma_2,(3)}\}$ are explicitly derived in Appendix~\ref{appendixb}. We will see in section~\ref{stochred} that the thirteen parameters above have degeneracies, therefore further reducing the number of free parameters of the theory.
		
		As has been pointed out for example in~\cite{Senatore:2014vja,Perko:2016puo}, the halo velocity divergence can be expanded to have a form similar to $\delta_h$. Indeed it is easy to see from equations (\ref{kernels1})-(\ref{kernels3}) and (\ref{eq:euler_bias_k}) that, neglecting stochastic and counter-terms, we can obtain $K_{\theta_h}^{(n)}$ by using the following choice of coefficients in (\ref{eq:euler_bias_k})
		\bea
		\label{velocity_coef}
		&&c^{(\theta_h)}_{\delta,1}(a) =1 \\ \nonumber
		&&c^{(\theta_h)}_{\alpha,(2)}(a) =\mG_1^\theta(a), \quad c^{(\theta_h)}_{\beta,(2)}(a) = \mG_2^\theta(a)\\ \nonumber
		&&c^{(\theta_h)}_{\alpha_\sigma,(3)}(a) =\mU_\sigma^\theta(a), \quad c^{(\theta_h)}_{\beta_\sigma,(3)}(a) = \mV_{\sigma 2}^\theta(a), \quad c^{(\theta_h)}_{\gamma_\sigma,(3)}(a) = \mV_{\sigma 1}^\theta(a)\\ \nonumber
		&&c^{(\theta_h)}_{\mathbb{I},(2)}(a)=c^{(\theta_h)}_{\alpha,(3)}(a)=c^{(\theta_h)}_{\beta,(3)}(a)=c^{(\theta_h)}_{\mathbb{I},(3)}(a)=0,
		\eea
		where again $\sigma \in \{1,2\}$.
		
		\subsection{Temporal degeneracies and a new functional form for the bias}\label{stochred}
		The formalism introduced in the previous section allows us to do bias expansions without the use of the EdS approximation. Of course in the appropriate limit, the bias expansions have to reduce to the expressions in the approximate case, which are described by only seven parameters. In this section, we will discuss the degeneracies that reduce the number of coefficients from thirteen to seven in both the approximate and exact case. However, as we will see, the functional form in the exact case slightly differs from the EdS approximated theory.
		
		From the explicit coefficients given in Appendix~\ref{appendixb} and the identities for the Green's function in Appendix~\ref{appendixa}, one can infer the following five relations
		\bea
		\label{eq:degen}
		&&c_{\alpha,(2)}+c_{\beta,(2)} = c_{\delta,1}\\ \nonumber
		&&c_{\alpha,(3)}+c_{\beta,(3)} = 2c_{\mathbb{I},(2)}\\ \nonumber
		&&c_{\beta_2,(3)}+c_{\alpha,(2)}-c_{\alpha_1,(3)}= \fr{1}{2}c_{\delta,1},\\ \nonumber
		&&c_{\alpha_1,(3)}+c_{\alpha_2,(3)}=c_{\gamma_1,(3)}+c_{\gamma_2,(3)}\\ \nonumber
		&&c_{\beta_1,(3)}+c_{\beta_2,(3)}+c_{\gamma_1,(3)}+c_{\gamma_2,(3)} =	\fr{1}{2}c_{\delta,1}, \nonumber
		\eea
		that hold without the EdS approximation. Furthermore, there is one relation that only holds with the EdS approximation $c_{\gamma_1,(3)}+c_{\beta_1,(3)}\stackrel{EdS}{=}\fr{3}{14}c_{\delta,1}$. We, therefore, define a function that parametrizes the departure from EdS
			\bea\label{Ydef}
			Y(a)\,c_{\delta,1}=-\fr{3}{14} c_{\delta,1}+c_{\gamma_1,(3)}+c_{\beta_1,(3)}.
			\eea
			Notice that $Y(a)$ is completely determined by functions that appear in (\ref{kernels3}) and a derivation can be found in Appendix~\ref{appendixb}. We get 
			\bea\label{Ynice}
			Y(a)=-\fr{3}{14}+\mV^{\delta}_{11}(a)+\mV^{\delta}_{12}(a).
			\eea
			In this form it is easy to see why $Y(a)\stackrel{EdS}{=}0$, since $\mV^{\delta}_{11}(a)\stackrel{EdS}{=}\fr{1}{6}$ and $\mV^{\delta}_{12}(a)\stackrel{EdS}{=}\fr{1}{21}$. Of course, since the EdS approximation is correct up to roughly percent level precision of the full power spectrum, we expect $Y(a)$ to be very small, and indeed it is zero in the matter-dominated era and increases to order $10^{-3}$ at late times as is shown in Figure~\ref{fig:relativeY}.
			
			 \begin{figure}[h!]
				\begin{center}
					\includegraphics[width=11cm]{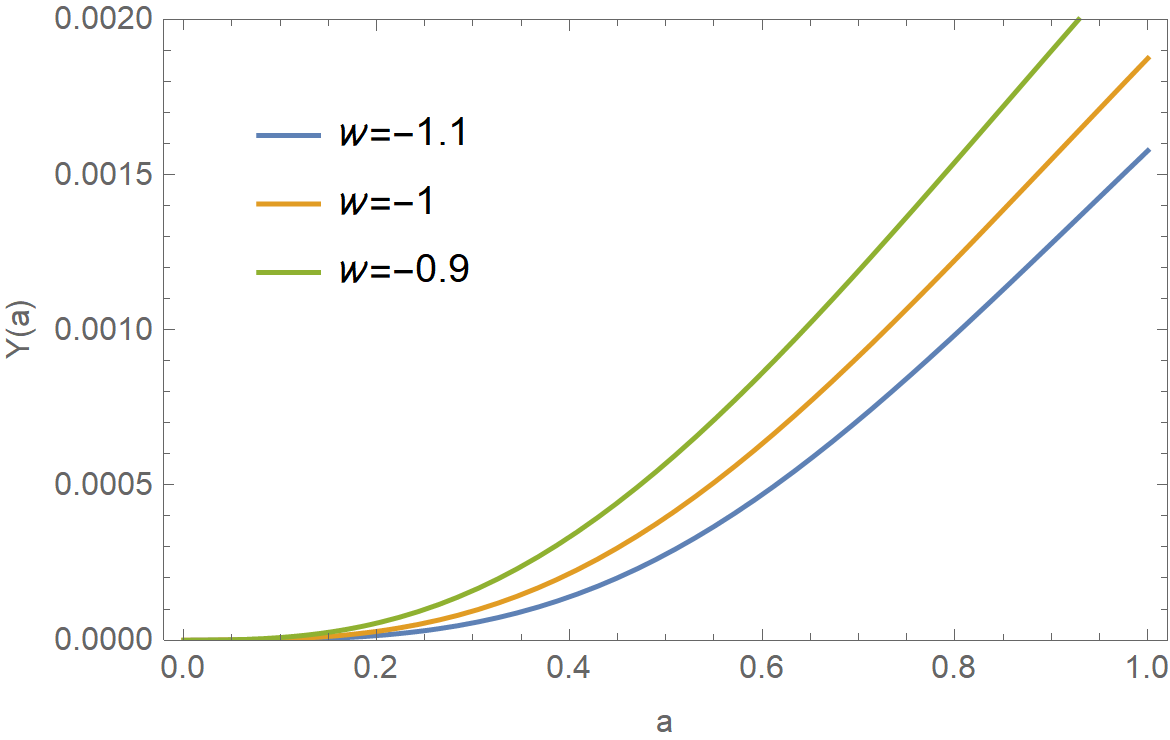}
					\caption{\label{fig:relativeY}\footnotesize Plot of the time evolution of the function $Y(a)$ {for different values of $w$}, that appears in the bias expansion for halos with exact time dependence. The departure from the EdS-approximation is proportional to $Y(a)$. Notice that the case of $w<-1$ and $c_s=1$ is unphysical~(see for example~\cite{Creminelli:2006xe,Creminelli:2008wc, Creminelli:2009mu}), but we plot it for illustration.}
				\end{center}
			\end{figure}
			
			Finally, we can rewrite equations (\ref{eq:euler_bias_k}) and (\ref{eq:kernelformh}) (still without stochastic and counter-terms) in terms of seven bias coefficients and the function $Y(a)$. To reduce the number of parameters} we replace the coefficients $c_{\beta,(2)}$, $c_{\beta,(3)}$, $c_{\alpha_1,(3)}$, $c_{\alpha_2,(3)}$, and $c_{\beta_2,(3)}$ through the identities in (\ref{eq:degen}) and remove $c_{\gamma_1,(3)}$ through the redefinition implied by (\ref{Ydef}). The expansion of the halo overdensity in Fourier space now reads
			\bea \label{eq:delta_h_CoI_t}
			\delta_h(\vec k,a)&=& c_{\delta,1}(a) \; \Big( \mathbb{C}^{(1)}_{\delta}(\vec k,a)+\mathbb{C}^{(2)}_{\delta}(\vec k,a)+\mathbb{C}^{(3)}_{\delta}(\vec k,a) +Y(a)\mathbb{C}^{(3)}_{Y}(\vec k,a)\Big)\\ \nonumber
			&+& c_{\alpha,(2)}(a) \; \Big(\mathbb{C}^{(2)}_{\alpha}(\vec k,a) +\mathbb{C}^{(3)}_{\alpha_1}(\vec k,a)\Big)\\ \nonumber &+&c_{\mathbb{I},(2)}(a) \;\Big(\mathbb{C}^{(2)}_{\mathbb{I}}(\vec k,a)+2\mathbb{C}^{(3)}_{\beta}(\vec k,a)\Big) \\ \nonumber
			&+& c_{\beta_1,(3)}(a) \; \mathbb{C}^{(3)}_{\beta_1}(\vec k,a)+ c_{\gamma_2,(3)}(a) \; \mathbb{C}^{(3)}_{\gamma_2}(\vec k,a)\\ \nonumber
			&+&c_{\alpha,(3)}(a) \; \mathbb{C}^{(3)}_{\alpha}(\vec k,a)+c_{\mathbb{I},(3)}(a) \mathbb{C}^{(3)}_{\mathbb{I}}(\vec k,a),
			\eea
		where the explicit $\mathbb{C}_i$ operators are given in Appendix~\ref{appendixb}. 
		
		In summary, we see that, at each time $a$, the field $\delta_h(\vec k,a) $ is obtained by the combination of seven functions, each one multiplied by an arbitrary bias coefficient. In particular, the six-dimensional space spanned by the functions appearing from the second to the last line of~(\ref{eq:delta_h_CoI_t}) is the same as a six-dimensional subspace spanned by the Basis of Descendants~(BoD) basis~\footnote{For completeness we give the transformation from the basis here (without $\mathbb{C}_Y$) to the BoD basis in Appendix~\ref{appendixd}.} as defined in EdS~\cite{Angulo:2015eqa}. Instead, the first function in (\ref{eq:delta_h_CoI_t}) is different than the one appearing in such EdS-defined BoD basis. At each time $a$, the part of this function that is of third order differs by the calculable, time- and $\vec k$-dependent function $Y(a)\mathbb{C}^{(3)}_{Y}(\vec k,a)$. Therefore, while at each time the space of functions spanned is still seven-dimensional, it is actually a different space. Of course, given that one can choose six of the seven basis functions to be the same as in the EdS-defined BoD basis, and given that $Y(a)$ is small, in practice the difference is not very large, as we will study later in section~\ref{secfits} (but, as we will also see there, not obviously negligibly small given the precision of upcoming experiments)~\footnote{Notice, that the correction proportional to $Y(a)$ would also be present if one were to impose that biased tracers depend on the long-wavelength fields in a local in time way, as done for example in~\cite{McDonald:2009dh}. While we recollect that this treatment is not justified by the time scales present in LSS (and a non-local in time treatment is instead necessary~\cite{Senatore:2014vja}), we here stress that the presence of the $Y(a)$ correction is just associated to the solution of the exact time dependence of the fields.}.
		
		Notice furthermore that, though the bias coefficients multiplying each function of the basis are incalculable within the EFT, they are in general different quantities once expressed in terms of the time kernels appearing in~(\ref{eq:euler_bias_new}) (as for example $c_\delta(a,a')$), with respect to the ones obtained in the EdS approximation. Therefore, if one had a theory that allowed to predict these time kernels, some of the resulting bias coefficients would be different in the two cases.


		
		\section{The halo-halo power spectrum in redshift space with exact time dependence}
		The basic formulas stated in this section were derived in~\cite{Senatore:2014vja} and used in~\cite{DAmico:2019fhj,Perko:2016puo}. We will briefly review the most important results that go into the halo-halo power spectrum in redshift space with exact time dependence. 
		\subsection{Halo bias with exact time dependence in redshift Space}
		The change from real space to redshift space, using the distant observer approximation, is just a change of coordinates
		\be
		\xv_r=\xv+\frac{\hat{z} \cdot \vec{v}}{a H}\hat{z} \ ,
		\ee
		where the $z$-axis was chosen to be along the line of sight. In Fourier space, the halo density perturbation changes under this coordinate transform into
		\be
		\delta_{h,r}(\kv,a)=\delta(\kv,a)+\int d^3x \ e^{-i \kv \cdot \xv} \left(\exp\left(- i \frac{k_z }{a H} v_{h,z}(\xv,a)\right)-1 \right) \left(1+\delta_h(\xv,a) \right) \ ,
		\label{rsdef}
		\ee
		where $\delta_{h,r}$ is the halo overdensity in redshift space.
		Following the procedure in~\cite{Senatore:2014vja,Perko:2016puo} we Taylor expand (\ref{rsdef}) in terms of the perturbations $\delta_h$ and $v_h$. There are products of operators in the Taylor expansion that are evaluated at the same location, which we have to renormalize by introducing new counter-terms. The halo bias expansion in redshift space, without the counter-terms, then becomes
		\bea\label{brackets}
		\delhr(\kv,a) &=&\delh(\kv,a)+ f_{+} \mu^2 \theta_h(\kv,a)\\ \nonumber 
		&&+i k\mu f_{+} \left [\dzdsq \theta_h \delh \right ]_{\kv}(a)-\frac{1}{2}k^2\mu^2 f_{+}^2 \left [\dzdsq \theta_h \dzdsq \theta_h \right]_{\kv}(a)\\ \nonumber
		&& -\frac{i}{6} k^3\mu^3 f_{+}^3 \left[ \dzdsq \theta_h \dzdsq \theta_h \dzdsq \theta_h \right]_{\kv}(a) -\frac{1}{2}k^2\mu^2 f_{+}^2 \left[ \dzdsq \theta_h \dzdsq \theta_h \delh \right]_{\kv}(a) + \ \ldots \ \ ,
		\eea
		where we have defined $\mu= k_z/k$ and $\ldots$ stands for the counter-terms and the stochastic terms, which we briefly discuss in section~\ref{countstoch}.
		We now perturbatively expand (\ref{brackets}) in terms of $\delta_h$ and $\theta_h$, to obtain the halo density perturbation in redshift space up to cubic order. Similar to equations (\ref{eq:kernelform}) and (\ref{eq:kernelformh}) we are interested in the halo integral kernels in redshift space $K^{(n)}_{h,r}(\q_1, \ldots , \q_n,a)$, which are defined at each order in perturbations by
		\be
		\delta^{(n)}_{h,r}(\kv,a)=\int \frac{d^3q_1}{(2\pi)^{3}}...\frac{d^3q_n}{(2\pi)^{3}}(2\pi)^{3}\delta_{D}(\vk-\vq_1-...-\vq_n)K^{(n)}_{h,r}(\q_1, \ldots , \q_n,a) \delta^{(1)}_{\vq_1}(a)...\delta^{(1)}_{\vq_n}(a) \ . 
		\ee
		The integrals of the form $[\ldots]_k$ in (\ref{brackets}) are given up to cubic order in Appendix~\ref{appendixe}. The explicit expressions for the full halo kernels in redshift space in terms of the halo kernels in real space from (\ref{eq:euler_bias_k}) and (\ref{velocity_coef}) read
		\bea
		\label{allredshift}
		K^{(1)}_{h,r} (\q_1,a) &=&K_{\delta_h}^{(1)} (\q_1,a)+f_{+} \mu^2 K_{\theta_h}^{(1)} (\q_1,a) \\ \nonumber
		K^{(2)}_{h,r}(\q_1, \q_2,a)&=&K_{\delta_h}^{(2)}(\q_1,\q_2,a)+f_{+} \mu^2 K_{\theta_h}^{(2)}(\q_1,\q_2,a) \\ \nonumber
		&&+\frac{1}{2} \mu f_{+} \left( \frac{ k q_{2z}}{q_2^2}+\frac{ k q_{1z}}{q_1^2} \right) K_{\theta_h} ^{(1)} (\q_1,a) K_{\delta_h} ^{(1)} (\q_2,a)\\ \nonumber
		&&+ \frac{1}{2}\mu^2f_{+}^2\left(\frac{k^2 q_{1z}q_{2z}}{q_1^2 q_2^2}\right)K_{\theta_h} ^{(1)} (\q_1,a) K_{\theta_h} ^{(1)} (\q_2,a) \\ \nonumber
		K^{(3)}_{h,r}(\q_1, \q_2, \q_3,a)&=&K_{\delta_h}^{(3)}(\q_1,\q_2,\q_3,a)+f_{+} \mu^2 K_{\theta_h}^{(3)}(\q_1,\q_2,\q_3,a) \\ \nonumber
		&&+ \mu f_{+} \left( \frac{k q_{3z}}{q_3^2} \right) K^{(2)}_{\delta_h} (\qv_1,\qv_2,a)K^{(1)}_{\theta_h} (\q_3,a) \\ \nonumber
		&&+\mu f_{+} \left( \frac{k(q_{1z}+ q_{2z})}{(\qv_1+\qv_2)^2} \right) K^{(2)}_{\theta_h} (\qv_1,\qv_2,a)K^{(1)}_{\delta_h} (\q_3,a)\ \\ \nonumber
		&& +\frac{1}{2}\mu^2 f_{+}^2 \left( \frac{k q_{1z}}{q_1^2}\frac{k q_{2z}}{q_2^2} \right) K^{(1)}_{\theta_h} (\q_1,a) K^{(1)}_{\theta_h} (\q_2,a) K^{(1)}_{\delta_h} (\q_3,a) \\ \nonumber
		&& + \mu^2 f_{+}^2 \left( \frac{k(q_{1z}+ q_{2z})}{(\qv_1+\qv_2)^2} \frac{k q_{3z}}{q_3^2} \right) K^{(2)}_{\theta_h} (\qv_1,\qv_2,a)K^{(1)}_{\theta_h} (\q_3,a) \\ \nonumber
		&&+\frac{1}{6}\mu^3 f_{+}^3 \left( \frac{k q_{1z}}{q_1^2}\frac{k q_{2z}}{q_2^2}\frac{k q_{3z}}{q_3^2} \right) K^{(1)}_{\theta_h} (\q_1,a) K^{(1)}_{\theta_h} (\q_2,a) K^{(1)}_{\theta_h} (\q_3,a) \ .
		\eea
		We are now able to write the full one-loop halo-halo power spectrum in redshift space. In terms of the halo kernels in redshift space from (\ref{allredshift}) it is given by 
		\bea
		\label{finalpsred}
		\langle \delhr(\kv,a) \delhr(\kv',a) \rangle' = \langle \delhr^{(1)}\delhr^{(1)} \rangle'+ \langle \delhr^{(2)} \delhr^{(2)} \rangle'+ 2 \langle \delhr^{(1)} \delhr^{(3)} \rangle' +\langle \delhr \delhr \rangle'_{\rm ct}+ \langle \delhr \delhr\rangle'_{\epsilon}\hspace{0.8in}\\ \nonumber
		=\bigl(K^{(1)}_{h,r}(a) \bigl)^2 P_{11}(k,a)+2 \int d^3 \q \ \left(K^{(2)}_{h,r}(\q,\kv-\q,a)_{\sym} \right)^2 P_{11}(|\kv-\q|,a)P_{11}(q,a)\hspace{0.3in} \\ \nonumber
		+6\int d^3 \q \ K^{(3)}_{h,r}(\q,-\q, \kv,a)_{\sym} K^{(1)}_{h,r}(a) P_{11}(q,a) P_{11}(k,a) + \langle \delhr \delhr \rangle'_{\rm ct}+ \langle \delhr \delhr \rangle'_{\epsilon}\ ,
		\eea
		where $P_{11}$ is the linear power spectrum and the contributions form counter-terms $\langle \delhr \delhr \rangle_{\rm ct}$ and stochastic terms $\langle \delhr \delhr \rangle_{\epsilon}$ are calculated in the next section. 
		Finally, we want to explicitly define the final bias parameters in terms of the coefficients in (\ref{eq:delta_h_CoI_t}). The halo kernels that enter into the power spectrum with the momentum dependence in
		(\ref{finalpsred}) read
		\begin{eqnarray}\label{finalredshiftkernel}
		K^{(1)}_{\delta_h}(a) &=&b_1 \\ \nonumber
		K^{(2)}_{\delta_h}(\vq,\vk-\vq,a)_{\sym} &=& \fr{b_1}{2 q}\fr{-k^2q+k^3x}{k^2+q^2-2kqx}+b_3\fr{ k^2 (1-x^2)}{k^2 + q^2 - 2 k q x}+b_2
		\\ \nonumber
		K^{(3)}_{\delta_h}(\vq,-\vq,\vk,a)_{UV-sub,\sym} &=&\fr{b_1}{42 q^2}\fr{-7 k^6 x^2 + k^2 q^4 (6 - 25 x^2 + 12 x^4) + 
			2 k^4 q^2 (3 - 10 x^2 + 14 x^4)}{(k^2 + q^2 - 2 k q x) (k^2 + q^2 + 2 k q x)} \\ \nonumber
		&&\quad+\fr{b_1\,Y(a)}{3}\fr{2k^4(1-x^2)+k^2q^2(2 - 6 x^2 + 4 x^4)}{(k^2 + q^2 - 2 k q x) (k^2 + q^2 + 2 k q x)},\\ \nonumber
		&&\qquad +\fr{b_{4}}{3}\fr{- 4k^4(1-2 x^2+x^4) -4k^2q^2 (1-2 x^2+x^4) }{(k^2 + q^2 - 2 k q x) (k^2 + q^2 + 2 k q x)},
		\end{eqnarray}
		where $x = \fr{\vq \cdot \vk}{qk}$ and we used the UV-subtracted third order kernel
		\be
		K^{(3)}_{\delta_h}(\vq,-\vq,\vk,a)_{UV-sub,\sym} = K^{(3)}_{\delta_h}(\vq,-\vq,\vk,a)_{\sym}-\lim_{\fr{q}{k} \to \infty } K^{(3)}_{\delta_h}(\vq,-\vq,\vk,a)_{\sym}.
		\ee
		Similarly for the $\langle \delhr^{(2)} \delhr^{(2)} \rangle$ contribution, we remove the UV-dependent part by subtracting $b_2^2\Sigma(a)^2$, where $\Sigma(a)^2 = \int\fr{d^3q}{(2\pi)^3}\left[P_{11}(q,a)\right]^2$.
		We can perform these shifts because we can absorb them into the counter-terms and stochastic terms.
		The number of coefficients we need reduces by three, as
		\bea
		{\mathbb{C}^{(3)}_{\mathbb{I}}}_{UV-sub,\sym}={\mathbb{C}^{(3)}_{\alpha}}_{UV-sub,\sym}={\mathbb{C}^{(3)}_{\beta}}_{UV-sub,\sym}={\mathbb{C}^{(3)}_{\alpha_1}}_{UV-sub,\sym}={\mathbb{C}^{(3)}_{\gamma_2}}_{UV-sub,\sym}=0.	
		\eea
		Explicitly, the final bias coefficients appearing in (\ref{finalredshiftkernel}) are given by
		\bea
		\label{eq:finalbias}
		b_1 = c_{\delta,1} \qquad b_2 = c_{\mathbb{I},(2)}\qquad
		b_3 = c_{\alpha,(2)}\qquad
		b_4 = c_{\beta_1,(3)}.
		\eea
		One can relate these coefficients to obtain the results in~\cite{Perko:2016puo} and we give the transformation in Appendix~\ref{appendixd}. In very close analogy, the halo velocity divergence kernels read
		\begin{eqnarray}\label{finalredshiftkernelth}
		K^{(1)}_{\theta_h}(a) &=& 1 \\ \nonumber
		K^{(2)}_{\theta_h}(\vq,\vk-\vq,a)_{\sym} &=&\fr{1}{2 q} \fr{-k^2q+k^3x}{(k^2+q^2-2kqx)}+\mG_1^{\theta}\fr{ k^2 (1-x^2)}{k^2 + q^2 - 2 k q x}
		\\ \nonumber
		K^{(3)}_{\theta_h}(\vq,-\vq,\vk,a)_{UV-sub,\sym} &=&\fr{1}{42 q^2}\fr{-7 k^6 x^2 + k^2 q^4 (6 - 25 x^2 + 12 x^4) + 
			2 k^4 q^2 (3 - 10 x^2 + 14 x^4)}{(k^2 + q^2 - 2 k q x) (k^2 + q^2 + 2 k q x)} \\ \nonumber
		&&\quad+\fr{Y(a)}{3}\fr{2k^4(1-x^2)+k^2q^2(2 - 6 x^2 + 4 x^4)}{(k^2 + q^2 - 2 k q x) (k^2 + q^2 + 2 k q x)}.\\ \nonumber
		&&\qquad+ \fr{\mV_{12}^{\theta}}{3}\fr{- 4k^4(1-2 x^2+x^4) -4k^2q^2 (1-2 x^2+x^4) }{(k^2 + q^2 - 2 k q x) (k^2 + q^2 + 2 k q x)}.
		\end{eqnarray}
		Note, that these are the same kernels as for the halo overdensity in (\ref{finalredshiftkernel}), but with different coefficients. This is essentially an extension of the identity in (\ref{velocity_coef}), which, after accounting for the degeneracies and the UV-subtraction, gives us
		\bea
		\label{eq:finalbiast}
		b_1^{(\theta_h)} = 1 \qquad 
		b_2^{(\theta_h)} =0\qquad
		b_3^{(\theta_h)} = \mG_1^{\theta}\qquad
		b_4^{(\theta_h)} = \mV_{12}^{\theta}.
		\eea
		\subsection{Counter-terms and stochastic halo bias}\label{countstoch}
		To complete the halo-halo power spectrum calculation, we now tend to the terms in (\ref{finalpsred}) that we have ignored so far. Namely the counter-terms from real and redshift space, as well as the stochastic terms.
		
		We start with the dark matter counter-terms that are in (\ref{eq:masterreal2}) and we neglected in (\ref{eq:master2}) and in their solution (\ref{deltaexpansion}). They stem from the non-local in time effective stress-tensor, which up to linear order in fields is given by
		\be
		\label{eq:stresstensor}
		\fr{1}{\rho}\d_j\tau^{ij} = \int \frac{da'}{a'^2 H(a')} K(a,a') \d^i \delta(a',\xfl(\vx,a,a'))\ +\ldots,
		\ee
		where $\rho$ is the background density and $K(a,a')$ is a time kernel. The effective stress tensor enters the velocity divergence equation (\ref{eq:master2}) at third order through $\fr{\d_i\d_j{\tau^{ij}}^{(1)}}{a\rho}$. Similar to (\ref{eq:coefs}) we can absorb the time integral into a coefficient, since at linear order the EdS approximation is exact and we can split the time dependence from the momentum dependence. 
		
		The resulting counter-terms for the halo kernels in redshift space read
		\bea
		\label{counterhalo}
		c_{\rm ct}^{(\delta_h)}\frac{k^2}{k_{\rm NL}^2}\delta^{(1)}+f_{+} \mu^2 c_{\rm ct}^{(\theta_h)}\frac{k^2}{k_{\rm NL}^2}\delta^{(1)},
		\eea 
		where $k_{\rm NL}$ is the wavenumber symbolizing the non-linear scale. Additionally, there are the counter-terms from the renormalization of the contact terms coming from (\ref{rsdef}) that we did not treat in (\ref{brackets}). They can be captured by two additional coefficients~\cite{Senatore:2014vja,Perko:2016puo}. Furthermore, we can absorb $c_{\rm ct}^{(\theta_h)}$ into one of these two additional coefficients and write the full counter-term in terms of three parameters
		\be
		\delhr^{(3, {\rm ct})}=c^{(\delta)}_{\rm ct} \frac{k^2}{k_{\rm NL}^2} \delta^{(1)}+\tilde c_{r,1}\mu^2 \left(\frac{k}{\km}\right)^2 \delta^{(1)}+ \tilde c_{r,2}\mu^4 \left(\frac{k}{\km}\right)^2 \delta^{(1)} \ .
		\ee
		The counter-terms enter the one-loop power spectrum as 
		\begin{align}
		&\langle \delhr(\kv,a) \delhr(\kv',a) \rangle_{\rm ct} = 2 \langle \delhr^{(1)}(\kv,a){\delhr^{(3,{\rm ct})}}(\kv',a) \rangle \\ \nonumber
		&\quad=2 P_{11}(k,a)(K^{(1)}_{\delta_h}(a)+f_{+} \mu^2K^{(1)}_{\theta_h} (a))\left( \mu^2 \left(\frac{k}{\km} \right)^2 \tilde{c}_{r,1}+ \mu^4 \left(\frac{k}{\km} \right)^2 \tilde{c}_{r,2} +c_{\rm ct}^{(\delta_h)} \left(\frac{k}{k_{\rm NL}} \right)^2 \right) 
		\ .
		\label{finalct}
		\end{align}
		
		We now move on to the stochastic terms that appear in the halo-halo power spectrum in redshift space, which are described by the stochastic field $\epsilon(\vx,a)$. It is assumed that the stochastic field only correlates with itself and the contribution is inversely proportional to the typical halo density $\langle \epsilon \epsilon \rangle \sim 1/\bar{n}$~\cite{Senatore:2014eva,Angulo:2015eqa}. 
		
		As was established in~\cite{Senatore:2014vja} and~\cite{Perko:2016puo}, the renormalized stochastic terms entering $\delta_{h,r}$ that come from the stochastic expansions of $\delta_h$ and $\theta_h$ are given by
		\be
		\delta_{h,r}^{(\epsilon)}={d}^2_{1,ren} \epsilon + {d}^2_{2,ren} \left( \frac{k}{\km} \right)^2\epsilon+ \ \ldots \ .
		\label{firststoch}
		\ee
		Additionally, there are stochastic terms $\delta_{stoch}$ that come from the renormalization of the contact terms in redshift space. $\delta_{stoch}$ can correlate with itself and with $\delta_{h,r}^{(\epsilon)}$. Finally the full stochastic contribution to the halo-halo power spectrum in redshift space, which includes both the real-space and redshift-space stochastic correlations, reads
		
		\be
		\langle \delta_{h,r}\delta_{h,r} \rangle_\epsilon =\frac{1}{\bar{n}} \left(c_{\epsilon,1} + c_{\epsilon,2} \left( \frac{k}{\km} \right)^2 +c_{\epsilon,3} f_{+}\mu^2 \left( \frac{k}{\km} \right)^2 \right) \ .
		\label{stochfin}
		\ee
		
		In conclusion, we need six coefficients $\{c^{(\delta)}_{\rm ct},\tilde c_{r,1},\tilde c_{r,2},c_{\epsilon,1},c_{\epsilon,2},c_{\epsilon,3}\}$ to account for the counter-terms and the stochastic contribution to the halo-halo power spectrum in redshift space. For more details see~\cite{Carrasco:2012cv,Senatore:2014vja,Angulo:2015eqa,Perko:2016puo}.
		
		%
		%
		%
		%
		
		\section{Comparisons with the EdS approximation} \label{secfits}
		Next, we want to compare the one-loop halo power spectrum in redshift space with exact time dependence, to the EdS approximated case. The formalism introduced in the previous sections applies to a generic $w$CDM cosmology. We here show the results only for $w=-1$, {\it i.e.} $\Lambda$CDM. The analogous results for $w$CDM are almost the same, simply differing by a relative factor of order $(1+w)\ll1$, so we avoid to explicitly present them since the conclusions do not change.
	 
	 Note, that there are two causes for the exact time dependence power spectrum to differ from the approximate one. In (\ref{allredshift}) and (\ref{finalpsred}) we see that the time dependence of the halo power spectrum in redshift space is captured by the overdensity and velocity divergence halo kernels in real space. From equation (\ref{finalredshiftkernel}) we get that the time dependence of the real-space halo overdensities depends on the incalculable bias coefficients $b_1(a)$, $b_2(a)$, $b_3(a)$, $b_4(a)$, as well as the calculable function $Y(a)$. However, the time dependence of the halo velocity divergence kernels in real space, given in (\ref{finalredshiftkernelth}), is fully determined by calculable functions. Therefore, to determine the impact of the EdS approximation on the redshift-space power spectrum for galaxies, we need to find an estimate for the time dependence of the bias coefficients, which is what we are going to do next.
		
		In~\cite{DAmico:2019fhj} the bias coefficients were measured using the EFTofLSS with EdS approximation. To estimate the value of these coefficients in the exact case, we compute $\Delta b_n = b_{n}-b_{n,\text{EdS}}$ for each of the bias coefficient, using the explicit definitions given in Appendix~\ref{appendixb} and their EdS approximations. It is easy to see from equations (\ref{eq:finalbias}), (\ref{third order final coefficients}) and (\ref{coefficientraw}) that $\Delta b_1 = 0$ and $\Delta b_2 = 0$. From the same equations, we get that
		\be
		\label{eq:db3}
		\Delta b_3(a) = \int^a\fr{da'}{a'}c_\delta(a,a')\fr{D_{+}(a')^2}{D_{+}(a)^2}\left(\mG_1^\delta(a')-\fr{5}{7}\right),
		\ee
		and a similar expression can be found for $\Delta b_4(a)$. 
		We can see that $\Delta b_3(a)$ and $\Delta b_4(a)$ depend on the time kernels, such as $c_\delta(a,a')$. Therefore, to estimate the change in the value of the bias coefficients with respect to the EdS approximation, we need an ansatz for $c_\delta(a,a')$.
		
		From Press-Schechter we have a rough estimate that $b_1(M,a) = 1-\fr{1}{\delta_c}+\fr{\delta_c}{\int^{1/M} d^3kP(a)}$, where $\delta_c\simeq1.7$ is the critical overdensity. If we include the time dependence of all the loop contributions into the power spectrum, integrated up to some mass scale, we can approximate $\int^{1/M} d^3k\; P(k,a)\sim a^2e^{-a^2}$. Therefore, at a fixed mass, the Press-Schechter formula for the bias now gives $b_1(a) \sim 1-\fr{1}{\delta_c}+c_g \fr{e^{a^2}}{a^2}$, where $c_g$ is a constant fit to the collapsed object of interest, such as halos or galaxies (in our case these are the coefficients measured for galaxies in~\cite{DAmico:2019fhj}). Now, since we are interested in biases of order one or larger at $a\lesssim1$, and $1-\fr{1}{\delta_c} \simeq 0.4$, and the term $c_g e^{a^2}/a^2$ increases as $a$ decreases, for the purpose of our estimate, we drop the $a$-independent term such that $b_1(a) \sim c_g \fr{e^{a^2}}{a^2}$. We thus define the kernel to be 
		\be \label{eq:greenestimate}
		c_\delta(a,a')= c_g \fr{1+2(a-a')^2}{a^2}e^{(a-a')^2},
		\ee
		such that from (\ref{eq:coefs}) and (\ref{eq:finalbias}) we approximately get $b_1(a)= b_1(a_{*})\fr{a_{*}^2}{a^2}e^{(a^2-a_{*}^2)}$, for some fixed time $a_{*}$~(\footnote{We chose this functional form because even though the exponential does not have a large quantitative impact ({\it i.e.} it could be dropped), it makes the evaluation of the time integrals easier.}).
		
\begin{figure}[h!]
			\begin{center}
				\includegraphics[width=12cm]{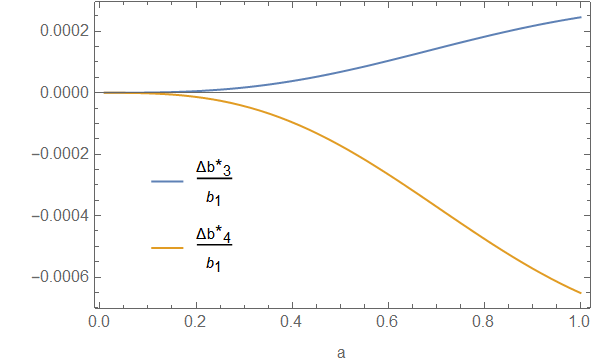}
				\caption{\label{fig:biasdif}\footnotesize Diagrammatic representation of $\Delta b_3^*(a)$ and $ \Delta b_4^*(a)$, relative to the linear bias $b_1(a)$ as a function of the scale factor. The functions, $\Delta b_3^*(a)$ and $ \Delta b_4^*(a)$ are an estimate for the change in the bias coefficients due to the EdS approximation.}
			\end{center}
		\end{figure}

		We denote the specific estimates, using equation (\ref{eq:greenestimate}) to calculate $\Delta b_3(a)$ and $\Delta b_4(a)$, by $\Delta b_3^*(a)$ and $\Delta b_4^*(a)$. These functions are depicted in Figure~\ref{fig:biasdif} relative to $b_1(a)$, where one can see that the EdS approximation gets worse as one moves forward in time, which was to be expected. However, relative to the linear bias $b_1$, $\Delta b_3^*(a)$ and $\Delta b_4^*(a)$ are of order $10^{-4}$.
		
		We now want to quantify the effect that the EdS approximation has on the one-loop halo power spectrum in redshift space. We here give plots for the effect in real space $P_{\rm Real}$, the monopole $P_{0}$ and the quadrupole $P_{2}$, all of which are resummed using IR-resummation~\cite{Senatore:2014via,Perko:2016puo} to correctly account for the BAO peaks\footnote{Notice that the IR-resummation is not affected by the inaccuracy of the EdS approximation.}. For the approximate cases ($P^{\text{EdS-approx}}_s$), where $s\in\{\rm Real,0,2\}$, we use the coefficients recently measured in~\cite{DAmico:2019fhj} (see Appendix~\ref{appendixf}), where the EdS approximation was used. In the exact cases ($P^{\text{Exact}}_s$), we rely on the future measurement of the bias coefficients. However, we can use the estimates from Figure~\ref{fig:biasdif} to here give three versions of plots, that illustrate the difference between the EdS approximation and the exact case. 
		
\begin{figure}[h!]
			\begin{center}
				\includegraphics[width=12cm]{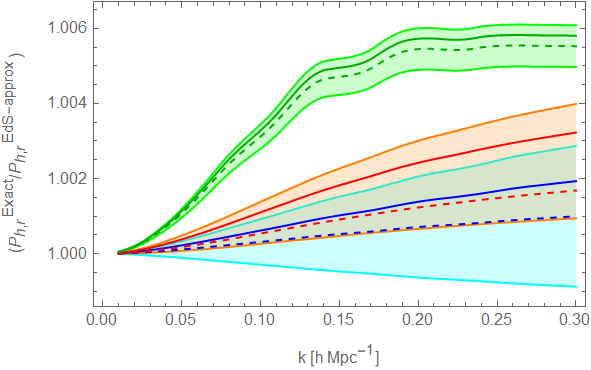}
				\caption{\label{fig:kevolution}\footnotesize Diagrammatic representation of the ratio of the exact galaxy power spectrum in redshift space over the approximate case as a function of $k$ at $a^* = 0.6$. The plots show the ratios of the real parts $P_{\rm Real}^{\text{Exact}}(k, a^*)/P_{\rm Real}^{\text{EdS-approx}}(k, a^*)$ (blue/cyan), the monopoles $P_{0}^{\text{Exact}}(k, a^*)/P_{0}^{\text{EdS-approx}}(k, a^*)$ (red/orange) and the quadrupoles $P_{2}^{\text{Exact}}(k, a^*)/P_{2}^{\text{EdS-approx}}(k, a^*)$ (dark/light green) of the galaxy power spectrum in redshift space. For the bias coefficients with EdS approximation we used $b_{1,\text{EdS}}(a^*) = 2.4$, $b_{2,\text{EdS}}(a^*) = -0.4$, $b_{3,\text{EdS}}(a^*) = 2.1$, $b_{4,\text{EdS}}(a^*) = 0$, $c_{ct,\text{EdS}}(a^*) = 0$, $\tilde{c}_{r,1,\text{EdS}}(a^*) = -8.6\left(\text{k}_\text{M}/\text{hMpc}^{-1}\right)^2$, $\tilde{c}_{r,2,\text{EdS}}(a^*) = 0,\quad c_{\epsilon,1,\text{EdS}}(a^*) = 1.4$ and $c_{\epsilon,2,\text{EdS}}(a^*) = -4.3\left(\text{k}_\text{M}/\text{hMpc}^{-1}\right)^2$ from~\cite{DAmico:2019fhj}. Furthermore, we have $Y(a^*) = 6*10^{-4}$. The dashed lines represent the effect of the approximation that comes from redshift space and the contribution multiplied by $Y(a^*)$ only {\it i.e.} $b_3(a^*) = b_{3,\text{EdS}}(a^*)$ and $ b_4(a^*) = b_{4,\text{EdS}}(a^*)$. The estimate from~(\ref{eq:greenestimate}), where $b_3(a^*) = b_{3,\text{EdS}}(a^*)+\Delta b_3^*(a^*)$ and $ b_4(a^*) = b_{4,\text{EdS}}(a^*)+\Delta b_4^*(a^*)$ (at $a^* = 0.6$ we have $\Delta b_3^*(a^*) = 2*10^{-4} $ and $\Delta b_4^*(a^*) = -6*10^{-4}$), is depicted by the darker solid lines. The lighter shaded areas are bounded from below ($-$) and above ($+$) by $b_3(a^*)= b_{3,\text{EdS}}(a^*)\pm2*\Delta b_3^*(a^*)$ and $b_4(a^*)= b_{4,\text{EdS}}(a^*)\pm2*\Delta b_4^*(a^*)$.}
			\end{center}
		\end{figure}

\begin{figure}[h!]
		\begin{center}
			\includegraphics[width=12cm]{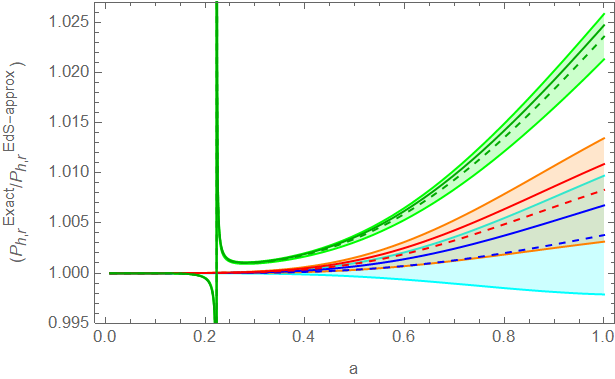}
			\caption{\label{fig:evolution}\footnotesize Diagrammatic representation of the ratio of the exact galaxy power spectrum in redshift space over the approximate case as a function of the scale factor at $k^* = 0.2\ \text{hMpc}^{-1}$. The plots show the ratios of the real parts $P_{\rm Real}^{\text{Exact}}(k^*, a)/P_{\rm Real}^{\text{EdS-approx}}(k^*, a)$ (blue/cyan), the monopoles $P_{0}^{\text{Exact}}(k^*,a)/P_{0}^{\text{EdS-approx}}(k^*,a)$ (red/orange) and the quadrupoles $P_{2}^{\text{Exact}}(k^*,a)/P_{2}^{\text{EdS-approx}}(k^*,a)$ (dark/light green) of the galaxy power spectrum in redshift space. For the bias coefficients with EdS approximation we used $b_{1,\text{EdS}}(a^*) = 2.2$, $b_{2,\text{EdS}}(a^*) = -0.4$, $b_{3,\text{EdS}}(a^*) = 1.9$, $b_{4,\text{EdS}}(a^*) = 0$, $c_{ct,\text{EdS}}(a^*) = 0$, $\tilde{c}_{r,1,\text{EdS}}(a^*) = -8\left(\text{k}_\text{M}/\text{hMpc}^{-1}\right)^2$, $\tilde{c}_{r,2,\text{EdS}}(a^*) = 0,\quad c_{\epsilon,1,\text{EdS}}(a^*) = 1.3$ and $c_{\epsilon,2,\text{EdS}}(a^*) = -4\left(\text{k}_\text{M}/\text{hMpc}^{-1}\right)^2$ from~\cite{DAmico:2019fhj} at $a^* = 0.64$. The coefficients were promoted to functions through the time dependence implied by (\ref{eq:greenestimate}). Furthermore, we use the calculable time dependence of $Y(a)$ from (\ref{Ynice}). The dashed lines represent the effect of the approximation that comes from redshift space and the contribution multiplied by $Y(a)$ only, {\it i.e.} $b_3(a) = b_{3,\text{EdS}}(a)$ and $ b_4(a) = b_{4,\text{EdS}}(a)$. The estimate from (\ref{eq:greenestimate}), where $b_3(a) = b_{3,\text{EdS}}(a)+\Delta b_3^*(a)$ and $ b_4(a) = b_{4,\text{EdS}}(a)+\Delta b_4^*(a)$ ($\Delta b_3^*(a)$ and $\Delta b_4^*(a)$ are shown in Figure~\ref{fig:biasdif}), is depicted by the darker solid lines. The lighter shaded areas are bounded from below ($-$) and above ($+$) by $b_3(a)= b_{3,\text{EdS}}(a)\pm2*\Delta b_3^*(a)$ and $b_4(a)= b_{4,\text{EdS}}(a)\pm2*\Delta b_4^*(a)$.}
		\end{center}
	\end{figure}

		First, we implement the estimate we did in (\ref{eq:greenestimate}), where the relative difference in the bias coefficients is of order $10^{-4}$, as given by Figure~\ref{fig:biasdif}. They are depicted in darker colored solid lines as a function of $k$ in Figure~\ref{fig:kevolution} and as a function of the scale factor in Figure~\ref{fig:evolution}. Next, we compute a conservative, but unambiguous estimate, which is to assume that the bias coefficients are not affected by the EdS approximation, {\it i.e.} $\Delta b_3 = 0$ and $\Delta b_4 = 0$. This essentially means that the effect is only due to the difference in the time dependences of $K_{\theta_{h}}^{(n)}$ from (\ref{finalredshiftkernelth}) and to the additional contribution that is multiplied by $Y(a)$ in (\ref{finalredshiftkernel}). This version of the plots is shown by the dashed lines in Figure~\ref{fig:kevolution} and Figure~\ref{fig:evolution}. Lastly, we give a band in which we expect the effect to lie in. The band is given by the EdS coefficients plus and minus two times the estimates from (\ref{eq:greenestimate}) that was considered in Figure~\ref{fig:biasdif}. It is depicted as the lightly shaded areas in Figure~\ref{fig:kevolution} and Figure~\ref{fig:evolution}.

	By looking at the quadrupole in Figure~\ref{fig:kevolution} and Figure~\ref{fig:evolution}, we see that the largest effect comes from the transformation into redshift space, and the estimate of the bias coefficients only dampens or enhances this effect. This is due to the fact that the EdS approximation is worse for the velocity divergence than for the density perturbation. Further checks with different coefficients and approximations can be found in Appendix~\ref{appendixf}, where depending on the size of the bias coefficients the effect can be up to a factor two larger.
	
	We can see from Figure~\ref{fig:kevolution} that the EdS approximation becomes more important at higher $k$. This is to be expected since at the linear level the EdS approximation is exact. Therefore, the EdS approximation only affects the loop terms, which become important only at higher $k$. Furthermore, from Figure~\ref{fig:evolution} we get the expected temporal evolution of the ratios of the power spectra. At early times ($a<0.1$), {\it i.e.} in the matter-dominated era, the EdS approximation is almost exact, and therefore we see that the ratios all stay at unity up to $a\simeq0.1$. However, at late times (for example $a = 0.85$ ($z = 0.18$)) the effect becomes quite large, even 1.7\% for the quadrupole and 0.8\% for the monopole. 
 
		In a last step, we want to discuss the applicability to data. Figure~\ref{fig:kevolution} and~\ref{fig:evolution} show that the physical difference between an exact time dependence and the approximate one is significant at late times. We here want to check if a change in the bias coefficients in the approximate case can account for this difference.
		
		For the analysis we take $a = 0.6$ ($z = 0.67$) like in Figure~\ref{fig:kevolution}. Furthermore, we use the galaxy power spectrum	in redshift space with the exact time dependences of $K_{\theta_{h}}^{(n)}$ and $Y(a)$, and fix the bias coefficients through the coefficients measured in~\cite{DAmico:2019fhj} plus the estimates $\Delta b_3^*$ {and} $\Delta b_4^*$ calculated using equation~(\ref{eq:greenestimate}). We then take the galaxy power spectrum in redshift space computed using the EdS approximation and use a best fit method to fit it to the exact time dependence galaxy power spectrum in redshift space. In this fit, we allow the biases that are expected to change between the exact time dependence and the EdS approximation to vary. As mentioned at the beginning of the section, since the time kernels such as $c_{\delta}(a,a')$ do not change due to the EdS approximation, only $b_3$ and $b_4$ can be affected by said approximation. We, therefore let $b_3$ and $b_4$ vary within $\pm 10^{-3}$, which is an order of magnitude larger than the estimated differences $\Delta b_3^*$ and $\Delta b_4^*$ (at $a=0.6$, $b_1$ is of order one)~\footnote{ \label{footnote:estimate} An alternative procedure would be to allow for all the bias coefficients to shift arbitrarily between the exact treatment and the EdS approximation. While such a procedure would show that the EdS-approximated predictions can fit the exact ones with much higher accuracy, we believe such a procedure would overemphasize the effectiveness of the EdS approximation. In fact, as mentioned, we expect the bias coefficients to differ due to the EdS approximation, relative to the linear bias, by about $10^{-4}$. If we were to allow the bias coefficients to vary in larger ranges, the coefficients may get unphysical. A consequence of this would probably be that the cosmological parameters that are extracted with this procedure would be systematically biased, even though the functional form of the predictions between the EdS approximated one and the exact one are very similar.
		
		In this regard, the situation is similar to the one we would encounter if we were to allow the bias coefficients to shift arbitrarily in order to fit the power spectrum of the observational data beyond where the one-loop approximation holds. Even though in this way a good fit could be obtained up to a higher wavenumber, the inferred cosmological parameters would be biased, as verified in~\cite{DAmico:2019fhj,Colas:2019ret}.	Indeed, we plan to explicitly quantify the effect of the EdS approximation directly in the extraction of the cosmological parameters in upcoming work. 
			}.

	Of course if we let $b_3$ and $b_4$ vary relatively by a factor of $10^{-3}$, we can, at least partially, absorb $\Delta b_3^*$ and $\Delta b_4^*$. However, we here want to check, how well a variation of $b_3$ and $b_4$ can absorb the changes due to the halo velocity divergence and the contribution from the $Y(a)$ term, that are depicted by the dashed lines in Figures~\ref{fig:kevolution} and~\ref{fig:evolution}. It is not obvious to what extent this is doable.
		
\begin{figure}[h!]
				\begin{center}
					\includegraphics[width=11.2cm]{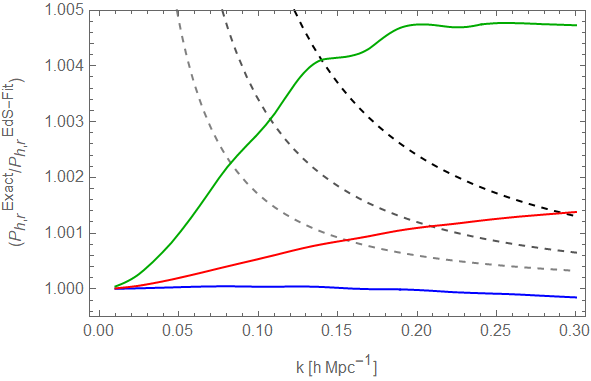}
					\caption{\label{fig:fit}\footnotesize The figure shows the ratio of the exact galaxy power spectrum in redshift space over a fit of the exact galaxy power spectrum in redshift space obtained by changing the bias coefficients in the EdS approximated case. The ratio is given as a function of $k$ at $a^* = 0.6$. The plots show the ratios of the real parts $P_{\rm Real}^{\text{Exact}}(k, a^*)/P_{\rm Real}^{\text{EdS-Fit}}(k, a^*)$ (blue), the monopoles $P_{0}^{\text{Exact}}(k, a^*)/P_{0}^{\text{EdS-Fit}}(k, a^*)$ (red) and the quadrupoles $P_{2}^{\text{Exact}}(k, a^*)/P_{2}^{\text{EdS-Fit}}(k, a^*)$ (dark green) of the galaxy power spectrum in redshift space. For the bias coefficients of the exact case we used the measured coefficients from~\cite{DAmico:2019fhj} and the estimate $\Delta b_3^*(a^*)$ and $\Delta b_4^*(a^*)$, as well as $Y(a^*)$, like in Figure \ref{fig:kevolution}. The best fit using the bias basis from the approximate EdS case gave us $\Delta b_{3,\text{EdS-Fit}}(a^*) =-2*10^{-3} $ and $\Delta b_{4,\text{EdS-Fit}}(a^*) =4*10^{-4} $. Furthermore the dashed lines are the expected error on the monopole, $1+\sigma(k)/4$, $1+\sigma(k)/2$ and $1+\sigma(k)$ for a survey like DESI~\cite{DESI}, where, very roughly, $\sigma(k)=0.024*\left(0.2\hinvMpc/k\right)^{3/2}$.} 
				\end{center}
			\end{figure}

		After this fitting procedure we obtain $P_s^{\text{EdS-Fit}}$, where $s\in\{0,2\}$ (and also $P_{\rm Real}^{\text{EdS-Fit}}$, which we will plot for consistency, though it is not observable), which is the EdS approximated galaxy power spectrum in redshift space, with a choice of bias coefficients (we call the resulting bias coefficients $b_{n,\text{EdS-Fit}}$ and define $\Delta b_{n,\text{EdS-Fit}} = b_n - b_{n,\text{EdS-Fit}}$) that best fits the exact case. The ratio of the two cases is depicted in Figure~\ref{fig:fit}. We see that, at $k=0.2\hinvMpc$, a change in the bias coefficients can account for the effect of the exact time dependence to a precision of $0.11\%$ for the monopole and $0.47\%$ for the quadrupole at $z = 0.67$, and, as suggested by Figure \ref{fig:evolution}, the magnitude of the effect most likely sharply increases at lower redshifts. One can compare this with the precision of future cosmological surveys such as DESI~\cite{DESI}, where we expect the error bars (given by the dashed lines in Figure \ref{fig:fit} for the monopole) to be, very roughly, 0.24\% for the monopole and 2.4\% for the quadrupole at $k\simeq 0.2\hinvMpc$. 
		
		As mentioned in footnote~\ref{footnote:estimate}, we expect the range we have chosen for the bias coefficients to be the appropriate one in order not to bias the information extracted from cosmological parameters. With the analysis provided here, we cannot be sure about this, and indeed if we let the biases vary in a larger range, the EdS approximated power spectrum would better fit the exact case. We plan to explicitly verify this in future work.		

		\section{Conclusion}
		In this paper, we remove the Einstein -- de Sitter approximation for biased tracers in redshift space in the EFTofLSS. We started with the bias expansion for collapsed objects treated with exact time dependence. We then further expanded the density perturbation and velocity divergence into a sum of momentum kernels, each one evolving with its own time dependence. Grouping together the momentum kernels of the biased expansion of the halos allows us to absorb temporal integrals into thirteen parameters, which can be further reduced to seven, by removing degeneracies among these parameters, just like in the EdS approximated solution. However, with respect to the EdS approximation, it is necessary to include an additional calculable time- and momentum-dependent contribution that is multiplied by the linear bias. Therefore, while biased tracers with exact time dependence can still be described by a set of seven bias parameters, the basis of the seven-dimensional vector space, in which the halo overdensities lie, is slightly different from the one present with the EdS approximation, and changes over time.
		
		The use of the exact time dependence for the density perturbations naturally introduced a basis for the momentum kernels that describes the halo overdensities. This is due to the fact that up to third order, the flow terms, as well as the tidal terms, can be expressed by the momentum kernels that appear in the density perturbations with exact time dependence. After accounting for temporal degeneracies, we are then automatically left with an irreducible basis for the biases.
				
		The coordinate transformation into redshift space with exact time dependence proceeds in a very similar way to~\cite{Perko:2016puo}. Neither the counter-terms nor the stochastic terms are affected up to cubic order.
		In total, the halo power spectrum in redshift space with exact time dependence is described by a set of ten coefficients (after UV-subtraction we have four bias coefficients and six counter-terms and stochastic terms). As mentioned before, with respect to the EdS approximation, there is an additional calculable contribution multiplying the linear bias, which appears as a consequence of the exact time dependence, and that already enters the halo power spectrum in real space.
		
		The quantitative effect of removing the EdS approximation on the galaxy power spectrum in redshift space is, as expected, larger than the one in real space. Since we computed the galaxy power spectrum in redshift space up to one-loop order, we stop the analysis in Figure~\ref{fig:kevolution} at $k = 0.3\unitsk$ and chose $k=0.2\unitsk$ in Figure~\ref{fig:evolution}, because the ratio of the power spectra might be affected more significantly by higher loop terms at higher $k$'s. In a survey such as BOSS (see for example~\cite{DAmico:2019fhj}), which is at $z = 0.57$, the error bars are, very roughly, $0.7\%$ for the monopole and $7\%$ for the quadrupole at $k\simeq 0.2\hinvMpc$. From Figure~\ref{fig:evolution} we get that at $z = 0.57$ the effect on the monopole is $0.3\%$ and $0.7\%$ for the quadrupole. At this level, the effect of the EdS approximation might, therefore, be almost negligible. Since we expect upcoming surveys, such as DESI~\cite{DESI}, to reduce the error bars to, very approximately, $0.24\%$ for the monopole and $2.4\%$ for the quadrupole, the exact time dependence might play a larger role at this level of precision. 
		
		By varying the bias coefficients in the EdS approximated case within a range that represented the physical deviation from the EdS approximation, we showed in Figure~\ref{fig:fit} that some of the change due to the exact time dependence can be absorbed into a small shift of the bias coefficients. This leads to a final effect of order $0.11\%$ in the monopole and $0.47\%$ in the quadrupole at $k=0.2\unitsk$. Therefore, the level of precision of the next generation of cosmological surveys is quantitatively similar to the impact the exact time dependence has on biased tracers in redshift space. The size of this effect depends on the allowed variation of the numerical value of the biases that we believe to be physically motivated. It would, therefore, be interesting to study more precisely the effect of the exact time dependence on realistic data and on the estimate of the cosmological parameters. While we leave this to future work, we point out that it appears to be not demanding to safely account for this effect using the formulas and implementation that we provide in this paper~\footnote{A Mathematica file can be found in the EFTofLSS code repository: \url{http://stanford.edu/~senatore/}}, for example by extending (without any significant slowdown) the publicly available code used for the BOSS analysis, such as~\cite{DAmico:2020kxu}.

		 \paragraph{Note Added:} While this paper was in advanced stage of completion, Ref.~\cite{Fujita:2020xtd} appeared, which finds the same conclusions to ours for the biased tracers in configuration space, {\it i.e.} for the results of Sec.~\ref{sec:biasedexact}, for $\Lambda$CDM and $w$CDM cosmologies.

		\section*{Acknowledgments}	
		YD is grateful for the kind
		hospitality at the Stanford Institute for Theoretical Physics~(SITP) at Stanford University and
		at the Kavli Institute for Particle Astrophysics and Cosmology~(KIPAC) at SLAC National
		Accelerator Laboratory. LS is partially supported by Simons
		Foundation Origins of the Universe program (Modern Inflationary Cosmology collaboration)
		and by NSF award 1720397.

		%
		%
		%
		%
		

		%
		%
		%
		%
		
		\begin{appendix}
			\section{Green's functions}\label{appendixa}
			At linear order, the time dependence is completely captured by the growths factor, which is defined as the solution of

			\be\label{delta-m34}
			\frac{d^2}{d\ln a^2}D(a)+\bigg(2+\frac{d\ln H}{d\ln a}\bigg)\frac{d}{d\ln a}D(a)-\fr{3}{2}\Omega_{m}(a)D(a)=0 \ ,
			\ee
			where 
			\be
			H(a) = H_0\sqrt{\Omega_{m,0}a^{-3}+\Omega_{D,0}a^{-3(1+w)}}
			\ee
			and we define the fractional matter and dark energy densities
			
			\be
			\Omega_m(a) = \Omega_{m,0}\fr{H_0^2}{H(a)^2}a^{-3} \quad\text{and}\quad \Omega_D(a) = \Omega_{D,0}\fr{H_0^2}{H(a)^2}a^{-3(1+w)},
			\ee
			in terms of their present day values $\Omega_{m,0}$ and $\Omega_{D,0}$.
			For generic $w$, the two solutions of (\ref{delta-m34}) are given in terms of the Hypergeometric functions~\cite{Lee:2009gb}. A growing mode
			\be
			D_{+}(a)=a\cdot\,{}_2F_1\left(\fr{w-1}{2w}, -\fr{1}{3w}, 
			1 -\fr{5}{6w} , -a^{-3 w} \fr{\Omega_{D,0}}{\Omega_{m,0}}\right)
			\ee
			and a decaying mode
			\be
			D_{-}(a)=a^{-\fr{3}{2}}\cdot\,{}_2F_1\left(\fr{1}{2w}, \fr{1}{2}+\fr{1}{3w}, 
			1 +\fr{5}{6w} , -a^{-3 w} \fr{\Omega_{D,0}}{\Omega_{m,0}}\right).
			\ee
			From there we get the linear growth indices $f_{\pm}\equiv \frac{d\ln D_{\pm}}{d\ln a}$. 
			
			In the special case where $w = -1$, i.e $\Lambda$CDM, the growing mode is\be
			D^{\Lambda}_{+}(a)=\frac52 \int^a_0\Omega^{\Lambda}_{m}(\ta)\frac{H^{\Lambda}(a)}{H^{\Lambda}(\ta)}d\ta,
			\ee
			and for the decaying mode we get
			\be
			D^{\Lambda}_{-}(a)=\frac{H^{\Lambda}(a)}{H_0\Omega_{m,0}^{1/2}} \ .
			\ee
			Furthermore the linear growth rates can be written as
			\be
			f^{\Lambda}_{+} ( a ) =\bigg(\frac52\frac{a}{D^{\Lambda}_{+} ( a ) }-\frac32\bigg)\Omega^{\Lambda}_{m}(a) \ ,
			\ee
			and
			\be
			f^{\Lambda}_{-}(a)=-\frac32\Omega^{\Lambda}_{m}(a) \ .
			\ee However, in what follows we will work with a generic value for $w$.
		
			To construct the solutions to the higher order time dependences appearing in (\ref{kernels1})-(\ref{kernels3}), coming from equations (\ref{eq:master1}) and (\ref{eq:master2}), we define the Green's functions
			\begin{align}
			&a \frac{d G^{\delta}_{\sigma}(a,\ta)}{da}-f_{+}(a)G^{\theta}_{\sigma}(a,\ta)=\lambda_{\sigma}\delta_D(a-\ta), \label{Green} \\
			&a \frac{d G^{\theta}_{\sigma}(a,\ta)}{da}-f_{+}(a)G^{\theta}_{\sigma}(a,\ta)+\fr{3}{2} \frac{\Omega_{m}}{f_{+}}\bigg(G^{\theta}_{\sigma}(a,\ta)-G^{\delta}_{\sigma}(a,\ta)\bigg)=(1-\lambda_{\sigma})\delta_D(a-\ta),
			\end{align}
			where $\lambda_1 = 1$ and $\lambda_2 = 0$.
			Explicitly the Green's functions are given by
			\begin{align}
			&G^{\delta}_1(a,\ta)=\frac{1}{\ta W(\ta)}\bigg(\frac{d D_{-}(\ta)}{d\ta}D_{+}(a)-\frac{d D_{+}(\ta)}{d\ta}D_{-}(a)\bigg) {\Theta}(a-\ta) \label{gdelta} \ ,\\
			&G^{\delta}_2(a,\ta)=\frac{f_{+}(\ta)/\ta^2}{W(\ta)}\bigg(D_{+}(\ta)D_{-}(a)-D_{-}(\ta)D_{+}(a)\bigg){\Theta}(a-\ta) \ , \\
			&G^{\theta}_1(a,\ta)=\frac{a/\ta}{f_{+}(a)W(\ta)}\bigg(\frac{d D_{-}(\ta)}{d\ta}\frac{d D_{+}(a)}{d a}-\frac{d D_{+}(\ta)}{d\ta}\frac{d D_{-}(a)}{d a}\bigg) {\Theta}(a-\ta) \ ,\\
			&G^{\theta}_2(a,\ta)=\frac{f_{+}(\ta)a/\ta^2}{f_{+}(a)W(\ta)}\bigg(D_{+}(\ta)\frac{d D_{-}(a)}{d a}-D_{-}(\ta)\frac{d D_{+}(a)}{d a}\bigg) {\Theta}(a-\ta) \ , \label{gtheta}
			\end{align}
			where $W(\ta)$ is the Wronskian of $D_+$ and $D_-$ 
			\be
			W(\ta)=\frac{dD_{-}(\ta)}{d\ta}D_{+}(\ta)-\frac{d D_{+}(\ta)}{d\ta}D_{-}(\ta) \ ,
			\ee
			$\Theta (a-\tilde a)$ is the Heaviside step function and we impose the boundary conditions \begin{align} \label{bound}
			& G^{\delta}_\sigma(a,\tilde a) = 0 \quad \quad \text{and} \quad\quad G^{\theta}_\sigma(a, \tilde a)=0 \quad \quad \text{for} \quad \quad \tilde a > a \ , \\
			&G^\delta_\sigma ( \tilde a , \tilde a ) = \frac{\lambda_\sigma}{\tilde a} \quad \hspace{.06in} \text{and} \hspace{.2in} \quad G^{\theta}_{\sigma} ( \tilde a , \tilde a ) = \frac{(1 - \lambda_\sigma)}{\tilde a} . \label{bound2}
			\end{align}
			Moving on we can define the time-dependent functions at second order
			
			\begin{align} \label{2nd-solution} 
			\mG^{\delta}_{\sigma}(a)&=\int^{1}_0 G^{\delta}_{\sigma}(a,\ta)\frac{f_{+}(\ta) D_{+}^2(\ta) }{ D_+^2 (a) }d\ta \ , \\
			\mG^{\theta}_{\sigma}(a)&=\int^{ 1}_0G^{\theta}_{\sigma}(a,\ta) \frac{f_{+}(\ta) D_{+}^2(\ta) }{ D_+^2 (a ) }d\ta, \label{2nd-solution2} 
			\end{align}
			for $\sigma = 1,2$. And then at third order
			
			\begin{align}\label{3rd-solution} 
			\mU^{\delta}_{\sigma}(a)=\int^{1}_0 G^{\delta}_{1}(a,\ta)\frac{f_{+}(\ta)D^3_{+}(\ta)}{ D^3_+ (a) }\mG^{\delta}_{\sigma}(\ta)d\ta,\\
			\mU^{\theta}_{\sigma}(a)=\int^{1}_0 G^{\theta}_{1}(a,\ta)\frac{f_{+}(\ta)D^3_{+}(\ta)}{ D^3_+ (a) }\mG^{\delta}_{\sigma}(\ta)d\ta,\\
			\mV^{\delta}_{\sigma\tilde\sigma}(a)=\int^{1}_0 G^{\delta}_{\tilde\sigma}(a,\ta)\frac{f_{+}(\ta)D^3_{+}(\ta)}{ D^3_+ (a) }\mG^{\theta}_{\sigma}(\ta)d\ta,\\
			\mV^{\theta}_{\sigma\tilde\sigma}(a)=\int^{1}_0 G^{\theta}_{\tilde\sigma}(a,\ta)\frac{f_{+}(\ta)D^3_{+}(\ta)}{ D^3_+ (a) }\mG^{\theta}_{\sigma}(\ta)d\ta.
			\label{3rd-solutionlast}
			\end{align}
			To derive the degeneracies pointed out in section~\ref{stochred} we need the following identities
			\bea \label{eq:timeidents}
			\mG_1^\delta+\mG_2^\delta=\mG_1^\theta+\mG_2^\theta&=&1\\ \nonumber
			\mV_{11}^\delta+\mV_{21}^\delta &=& \mU_1^\delta+ \mU_2^\delta\\ \nonumber
			\mV_{11}^\theta+\mV_{21}^\theta &=& \mU_1^\theta+ \mU_2^\theta\\ \nonumber
			\mV_{\sigma1}^\delta+\mV_{\sigma2}^\delta &=&\mV_{\sigma1}^\theta+\mV_{\sigma2}^\theta \\ \nonumber
			\mV_{11}^\delta+\mV_{21}^\delta+\mV_{12}^\delta+\mV_{22}^\delta&=&\frac{1}{2}\\ \nonumber
			\mV_{11}^\theta+\mV_{21}^\theta+\mV_{12}^\theta+\mV_{22}^\theta &=& \frac{1}{2}\\ \nonumber
			\mV_{\sigma1}^\delta+\mV_{\sigma2}^\delta &=& \mG_\sigma^{\delta} - \fr{\lambda_{\sigma}}{2}\\ \nonumber
			\mU^{\delta}_{1}+\mV^{\delta}_{21}&=&\fr{1}{2}\\ \nonumber
			\mU^{\theta}_{1}+\mV^{\theta}_{21}&=&\fr{1}{2}+\mG^{\theta}_{1}-\mG^{\delta}_{1},
			\eea
			where we remind that $\sigma\in \{1,2\}$. These relations can be inferred from equations (\ref{2nd-solution})-(\ref{3rd-solutionlast}) once one realizes the following
			\bea G^{\delta}_1(a,\ta)+G^{\delta}_2(a,\ta)&=&G^{\theta}_1(a,\ta)+G^{\theta}_2(a,\ta) = \fr{D_{+}(a)}{\ta D_{+}(\ta)}\Theta(a-\ta)\\
			G^{\delta}_1(a,\ta)-G^{\theta}_1(a,\ta)&=& \frac{W(a)}{\ta W(\ta)}\fr{D_+'(\ta)}{D_+'(a)} {\Theta}(a-\ta) . 
			\eea
			
			Furthermore, for the derivation of the functional form of $Y(a)$ in (\ref{Ynice}), it is important to note that 
				\bea\label{tintegral}
				\mV^{\delta}_{\sigma1}(a)+\mV^{\delta}_{\sigma2}(a)=\int^{a}_0 \frac{D'_{+}(\ta)D_{+}(\ta)}{D^2_+ (a) }\mG^{\theta}_{\sigma}(\ta)d\ta
			\eea
		 which is used in Appendix~\ref{appendixb}.
			\section{Halo kernels and degeneracy of halo bias parameters}\label{appendixb}
			The six kernels we will use throughout this section are defines as
			\begin{align}\label{alphakernels}
			&\alpha^1(\vq_1,\vq_2,\vq_3)=\alpha(\vq_3,\vq_1+\vq_2)\alpha_s(\vq_1,\vq_2), \\ &\alpha^2(\vq_1,\vq_2,\vq_3)=\alpha(\vq_3,\vq_1+\vq_2)\beta(\vq_1,\vq_2), \\
			&\beta^1(\vq_1,\vq_2,\vq_3)=2\beta(\vq_3,\vq_1+\vq_2)\alpha_s(\vq_1,\vq_2), \\ &\beta^2(\vq_1,\vq_2,\vq_3)=2\beta(\vq_3,\vq_1+\vq_2)\beta(\vq_1,\vq_2), \\
			&\gamma^1(\vq_1,\vq_2,\vq_3)=\alpha(\vq_1+\vq_2,\vq_3)\alpha_s(\vq_1,\vq_2), \\ &\gamma^2(\vq_1,\vq_2,\vq_3)=\alpha(\vq_1+\vq_2,\vq_3)\beta(\vq_1,\vq_2).
			\end{align}
			From equation (\ref{eq:euler_bias_new}) we get the following expressions in the bias expansion
			\begingroup
			\allowdisplaybreaks
			\bea\label{eq:euler_bias_kints}
			^*[\d_i \delta^{(1)}\; \sfrac{\d^i}{\d^2}\theta^{(1)}]&=&\alpha(\vq_1,\vq_2)-1\\ \nonumber
			^*[\d_i \delta^{(2)}(a')\;\sfrac{\d^i}{\d^2}\theta^{(1)}] &=&\sfrac{D_{+}(a')^2}{D_{+}(a)^2}\left(\mG_\sigma^\delta(a')\alpha^\sigma(\vq_1,\vq_2,\vq_3)-\mG_1^\delta(a')\alpha(\vq_1,\vq_2)-\mG_2^\delta(a')\beta(\vq_1,\vq_2)\right)\\ \nonumber
			^*[\d_i \delta^{(1)}\;\sfrac{\d^i}{\d^2}\theta^{(2)}(a'')] &=&\sfrac{D_{+}(a'')^2}{D_{+}(a)^2}\left(\mG_\sigma^\theta(a'')\gamma^\sigma(\vq_1,\vq_2,\vq_3)-\mG_1^\theta(a'')\alpha(\vq_1,\vq_2)-\mG_2^\theta(a'')\beta(\vq_1,\vq_2)\right)\\ \nonumber
			^* [\d_i \delta^{(1)}\; \sfrac{\d_j\d^i}{\d^2}\theta^{(1)}\; \sfrac{\d^j}{\d^2}\theta^{(1)}]&+&^*[\d_i\d_j \delta^{(1)}\;\sfrac{\d^i}{\d^2}\theta^{(1)}\sfrac{\d^j}{\d^2}\theta^{(1)}] =\alpha_1(\vq_1,\vq_2,\vq_3)-3\alpha(\vq_1,\vq_2)+2 \\ \nonumber
			^*[\delta^{(1)}\delta^{(2)}(a')]&=&\sfrac{D_{+}(a')^2}{D_{+}(a)^2}\left(\mG_1^\delta(a')\alpha(\vq_1,\vq_2)+\mG_2^\delta(a')\beta(\vq_1,\vq_2)\right)\\ \nonumber
			^*[\delta^{(1)} \d_i\delta^{(1)}\sfrac{\d^i}{\d^2}\theta^{(1)}]&=&\alpha(\vq_1,\vq_2)-1 \\ \nonumber
			^*[s^2]^{(2)}&=&\beta(\vq_1,\vq_2)-\alpha(\vq_1,\vq_2)+\sfrac{2}{3}\\ \nonumber
			^*[\sfrac{\d_i\d_j}{\d^2}\delta^{(2)}(a')\sfrac{\d^i\d^j}{\d^2}\delta^{(1)}]&-&\sfrac{1}{3} {^*}[\delta^{(2)}(a')\delta^{(1)}]=\\ \nonumber &=&\sfrac{D_{+}(a')^2}{D_{+}(a)^2}\bigg(\sfrac{1}{2}\mG_\sigma^\delta(a')\left(-\alpha^\sigma(\vq_1,\vq_2,\vq_3)+\beta^\sigma(\vq_1,\vq_2,\vq_3)-\gamma^\sigma(\vq_1,\vq_2,\vq_3)\right)\\ \nonumber
			&&\quad+\sfrac{2}{3}\mG_1^\delta(a')\alpha(\vq_1,\vq_2)+\sfrac{2}{3}\mG_2^\delta(a')\beta(\vq_1,\vq_2)\bigg) \\ \nonumber
			^*[\sfrac{\d_i\d_j}{\d^2}\delta^{(1)}\sfrac{\d^i\d^j}{\d^2}\theta^{(2)}(a')]&-&\sfrac{1}{3} {^*}[\delta^{(1)}\theta^{(2)}(a')]=\\ \nonumber &=&\sfrac{D_{+}(a')^2}{D_{+}(a)^2}\bigg(\sfrac{1}{2}\mG_\sigma^\theta(a')\left(-\alpha^\sigma(\vq_1,\vq_2,\vq_3)+\beta^\sigma(\vq_1,\vq_2,\vq_3)-\gamma^\sigma(\vq_1,\vq_2,\vq_3)\right)\\ \nonumber
			&&\quad+\sfrac{2}{3}\mG_1^\theta(a')\alpha(\vq_1,\vq_2)+\sfrac{2}{3}\mG_2^\theta(a')\beta(\vq_1,\vq_2)\bigg) \\ \nonumber
			^*[s_{lm}^{(1)}\d_i (s^{lm})^{(1)}\sfrac{\d^i}{\d^2}\theta^{(1)}]&=&\sfrac{1}{2}\left(-\alpha_1(\vq_1,\vq_2,\vq_3)+\alpha_2(\vq_1,\vq_2,\vq_3)+\sfrac{7}{3}\alpha(\vq_1,\vq_2)-\beta(\vq_1,\vq_2)-\sfrac{4}{3}\right)\\ \nonumber
			^*[\psi^{(3)}](a') &=& ^*[\eta^{(3)}](a')-\sfrac{D_{+}(a')^3}{D_{+}(a)^3}\left(\mG^{\delta}_{1}(a')-\mG^{\theta}_{1}(a')\right)\mG_\sigma^\delta(a')\big(-\alpha^\sigma(\vq_1,\vq_2,\vq_3)\\ \nonumber
			&&\quad+\beta^\sigma(\vq_1,\vq_2,\vq_3)-\gamma^\sigma(\vq_1,\vq_2,\vq_3)\big) \\ \nonumber
			^*[\delta^3]^{(3)} &=& 1\\ \nonumber
			^*[\delta s^2]^{(3)} &=&\beta(\vq_1,\vq_2)-\alpha(\vq_1,\vq_2)+\sfrac{2}{3} \\ \nonumber
			^*[s^3]^{(3)} &=& \sfrac{1}{2}\big(\alpha_1(\vq_1,\vq_2,\vq_3)-2\alpha_2(\vq_1,\vq_2,\vq_3)+\beta_2(\vq_1,\vq_2,\vq_3)-\gamma_2(\vq_1,\vq_2,\vq_3)\\ \nonumber
			&&\quad-\alpha(\vq_1,\vq_2) + \beta(\vq_1,\vq_2) + 4/9\big),
			\eea
			\endgroup
			where repeated $\sigma \in \{1,2\}$ are summed over and we used the notation, 
			\bea
			[X^{(n)}]_{\vk} &=& \int\frac{d^3q_1}{(2\pi)^{3}}\ldots\frac{d^3q_n}{(2\pi)^{3}}(2\pi)^{3}\delta_{D}(\vk-\vq_1-\ldots-\vq_n) \ {}^*[X^{(n)}] \delta^{(1)}_{\vq_1}(a)\ldots\delta^{(1)}_{\vq_n}(a).
			\eea
			Similarly to the second and third order density perturbation and velocity divergence, all of the expressions in (\ref{eq:euler_bias_kints}) only depend on the nine kernels $\{1,\alpha,\beta,\alpha_1,\alpha_2,\beta_1,\beta_2,\gamma_1,\gamma_2\}$. We can then factor out these momentum kernels and redefine the temporal coefficients to obtain the parameters that appear in (\ref{eq:euler_bias_k}) and (\ref{eq:delta_h_CoI_t})
			\begin{align}\label{third order final coefficients}
			&c_{\alpha,(2)}=c_{\delta_2,\mG_1^\delta}+c_{\delta,12}-c_{s^2,1}\\ \nonumber
			&c_{\beta,(2)}=c_{\delta_2,\mG_2^\delta}+c_{s^2,1}\\ \nonumber
			&c_{\mathbb{I},(2)} = -c_{\delta,12}+c_{\delta^2,1}+\sfrac{2}{3}c_{s^2,1}\\ \nonumber
			&c_{\alpha_1,(3)} =c_{\delta,\mU^{\delta}_{1}}+c_{\delta,\mG^{\delta}_{1}}+c_{\delta,{123}}-c_{s^2,\mG^{\delta}_{1}}-c_{s^2,{12}}-\sfrac{1}{2}\left(c_{st,\mG^{\theta}_{1}}-c_{st,\mG^{\delta}_{1}}\right)+c_{\psi,\mU^{\theta}_{1}}-c_{\psi,\mU^{\delta}_{1}}+ c_{\psi,\mG^{\delta}_{1}}+\sfrac{1}{2}c_{s^3}\\ \nonumber
			&c_{\alpha_2,(3)} =c_{\delta,\mU^{\delta}_{2}}+c_{\delta,\mG^{\delta}_{2}}-c_{s^2,\mG^{\delta}_{2}}+c_{s^2,{12}}-\sfrac{1}{2}\left(c_{st,\mG^{\theta}_{2}}-c_{st,\mG^{\delta}_{2}}\right)+c_{\psi,\mU^{\theta}_{2}}-c_{\psi,\mU^{\delta}_{2}}+ c_{\psi,\mG^{\delta}_{2}}-c_{s^3}\\ \nonumber
			&c_{\beta_1,(3)} =c_{\delta,\mV^{\delta}_{12}}+c_{s^2,\mG^{\delta}_{1}}+\sfrac{1}{2}\left(c_{st,\mG^{\theta}_{1}}-c_{st,\mG^{\delta}_{1}}\right)+c_{\psi,\mV^{\theta}_{12}}-c_{\psi,\mV^{\delta}_{12}}- c_{\psi,\mG^{\delta}_{1}}\\ \nonumber
			&c_{\beta_2,(3)} =c_{\delta,\mV^{\delta}_{22}}+c_{s^2,\mG^{\delta}_{2}}+\sfrac{1}{2}\left(c_{st,\mG^{\theta}_{2}}-c_{st,\mG^{\delta}_{2}}\right)+c_{\psi,\mV^{\theta}_{22}}-c_{\psi,\mV^{\delta}_{22}}- c_{\psi,\mG^{\delta}_{2}}+\sfrac{1}{2}c_{s^3}\\ \nonumber
			&c_{\gamma_1,(3)} =c_{\delta,\mV^{\delta}_{11}}+c_{\delta,\mG^{\theta}_{1}}-c_{s^2,\mG^{\delta}_{1}}-\sfrac{1}{2}\left(c_{st,\mG^{\theta}_{1}}-c_{st,\mG^{\delta}_{1}}\right)+c_{\psi,\mV^{\theta}_{11}}-c_{\psi,\mV^{\delta}_{11}}+ c_{\psi,\mG^{\delta}_{1}}\\ \nonumber
			&c_{\gamma_2,(3)} =c_{\delta,\mV^{\delta}_{21}}+c_{\delta,\mG^{\theta}_{2}}-c_{s^2,\mG^{\delta}_{2}}-\sfrac{1}{2}\left(c_{st,\mG^{\theta}_{2}}-c_{st,\mG^{\delta}_{2}}\right)+c_{\psi,\mV^{\theta}_{21}}-c_{\psi,\mV^{\delta}_{21}}+ c_{\psi,\mG^{\delta}_{2}}-\sfrac{1}{2}c_{s^3}\\ \nonumber
			&c_{\alpha,(3)} = -c_{\delta,\mG^{\delta}_{1}}-c_{\delta,\mG^{\theta}_{1}}-3c_{\delta,{123}}+2c_{\delta^2,\mG^{\delta}_{1}}+2c_{\delta^2,{12}}+\sfrac{4}{3}c_{s^2,\mG^{\delta}_{1}}+\sfrac{7}{3}c_{s^2,{12}}+\sfrac{2}{3}\left(c_{st,\mG^{\theta}_{1}}-c_{st,\mG^{\delta}_{1}}\right)-c_{\delta s^2}-\sfrac{1}{2}c_{s^3}\\ \nonumber
			&c_{\beta,(3)} = -c_{\delta,\mG^{\delta}_{2}}-c_{\delta,\mG^{\theta}_{2}}+2c_{\delta^2,\mG^{\delta}_{2}}+\sfrac{4}{3}c_{s^2,\mG^{\delta}_{2}}-c_{s^2,{12}}+\sfrac{2}{3}\left(c_{st,\mG^{\theta}_{2}}-c_{st,\mG^{\delta}_{2}}\right)+c_{\delta s^2}+\sfrac{1}{2}c_{s^3}\\ \nonumber
			&c_{\mathbb{I},(3)} = -2c_{\delta^2,{12}}+c_{\delta^3}-\sfrac{4}{3}c_{s^2,{12}}+\sfrac{2}{3}c_{\delta s^2}+2c_{\delta,{123}}+\sfrac{2}{9}c_{s^3},
			\end{align} 
			where in an intermediary step we defined the symbolic integrals over the time-dependent functions discussed in Appendix~\ref{appendixa}. They are given 
			\begin{alignat}{3}
			\label{coefficientraw}
			&\text{at first}&&\text{order by}\\ \nonumber
			&c_{\delta,1}(a) &&= \int^a \frac{da'}{a'}c_{\delta}(a,a')\frac{D_{+}(a')}{D_{+}(a)},\\ \nonumber
			&\text{at secon}&&\text{d order by}\\ \nonumber
			&c_{\delta_2,\mG^{\delta}_{\sigma}}(a) &&= \int^a \frac{da'}{a'}c_\delta(a,a')\frac{D_{+}(a')^2}{D_{+}(a)^2}\mG^{\delta}_{\sigma}(a') 
			&&c_{\delta,12}(a) = \int^a \frac{da'}{a'}c_\delta(a,a')\left[ \frac{D_{+}(a')}{D_{+}(a)}-\frac{D_{+}(a')^2}{D_{+}(a)^2}\right]\\ \nonumber
			&c_{s^2,1}(a) &&= \int^a \frac{da'}{a'}c_{s^2}(a,a') \frac{D_{+}(a')^2}{D_{+}(a)^2}
			&&c_{\delta^2,1}(a) = \int^a \frac{da'}{a'}c_{\delta^2}(a,a') \frac{D_{+}(a')^2}{D_{+}(a)^2},\\ \nonumber
			&\text{and at t}&&\text{hird order by}\\ \nonumber
			&c_{s^3}(a) &&= \int^a \frac{da'}{a'}c_{s^3}(a,a')\frac{D_{+}(a')^3}{D_{+}(a)^3}
			&&c_{\delta^3}(a) = \int^a \frac{da'}{a'}c_{\delta^3}(a,a')\frac{D_{+}(a')^3}{D_{+}(a)^3}\\ \nonumber
			&c_{\delta,\mU^{\delta}_{\sigma}}(a) &&= \int^a \frac{da'}{a'}c_\delta(a,a')\frac{D_{+}(a')^3}{D_{+}(a)^3}\mU^{\delta}_{\sigma}(a') 
			&&c_{\delta,\mG^{\delta}_{\sigma}}(a) = \int^a \frac{da'}{a'}c_\delta(a,a')\left[ \frac{D_{+}(a')^2}{D_{+}(a)^2}-\frac{D_{+}(a')^3}{D_{+}(a)^3}\right]\mG^{\delta}_{\sigma}(a')\\ \nonumber
			&c_{\delta,\mV^{\delta}_{\sigma \tilde{\sigma}}}(a) &&= \int^a \frac{da'}{a'}c_\delta(a,a')\frac{D_{+}(a')^3}{D_{+}(a)^3}\mV^{\delta}_{\sigma \tilde{\sigma}}(a')
			&&c_{\delta s^2}(a) = \int^a \frac{da'}{a'}c_{\delta s^2}(a,a')\frac{D_{+}(a')^3}{D_{+}(a)^3}\\ \nonumber
			&c_{\delta^2,\mG^{\delta}_{\sigma}}(a) &&= \int^a \frac{da'}{a'}c_{\delta^2}(a,a')\frac{D_{+}(a')^3}{D_{+}(a)^3}\mG^{\delta}_{\sigma}(a') 
			&&c_{\delta^2,{12}}(a) = \int^a \frac{da'}{a'}c_{\delta^2}(a,a')\left[ \frac{D_{+}(a')^2}{D_{+}(a)^2}-\frac{D_{+}(a')^3}{D_{+}(a)^3}\right]\\ \nonumber
			&c_{s^2,\mG^{\delta}_{\sigma}}(a) &&= \int^a \frac{da'}{a'}c_{s^2}(a,a')\frac{D_{+}(a')^3}{D_{+}(a)^3}\mG^{\delta}_{\sigma}(a') &&c_{s^2,{12}}(a) = \int^a \frac{da'}{a'}c_{s^2}(a,a')\left[ \frac{D_{+}(a')^2}{D_{+}(a)^2}-\frac{D_{+}(a')^3}{D_{+}(a)^3}\right]\\ \nonumber
			&c_{st,\mG^{\delta}_{\sigma}}(a) &&= \int^a \frac{da'}{a'}c_{st}(a,a')\frac{D_{+}(a')^3}{D_{+}(a)^3}\mG^{\delta}_{\sigma}(a') &&c_{st,\mG^{\theta}_{\sigma}}(a) = \int^a \frac{da'}{a'}c_{st}(a,a')\frac{D_{+}(a')^3}{D_{+}(a)^3}\mG^{\theta}_{\sigma}(a')\\ \nonumber
			&c_{\psi,\mU^{\delta}_{\sigma}}(a) &&= \int^a \frac{da'}{a'}c_\psi(a,a')\frac{D_{+}(a')^3}{D_{+}(a)^3}\mU^{\delta}_{\sigma}(a') 
			&&c_{\psi,\mV^{\delta}_{\sigma \tilde{\sigma}}}(a) = \int^a \frac{da'}{a'}c_\psi(a,a')\frac{D_{+}(a')^3}{D_{+}(a)^3}\mV^{\delta}_{\sigma \tilde{\sigma}}(a')\\ \nonumber
			&c_{\psi,\mU^{\theta}_{\sigma}}(a) &&= \int^a \frac{da'}{a'}c_\psi(a,a')\frac{D_{+}(a')^3}{D_{+}(a)^3}\mU^{\theta}_{\sigma}(a') 
			&&c_{\psi,\mV^{\theta}_{\sigma \tilde{\sigma}}}(a) = \int^a \frac{da'}{a'}c_\psi(a,a')\frac{D_{+}(a')^3}{D_{+}(a)^3}\mV^{\theta}_{\sigma \tilde{\sigma}}(a')\\ \nonumber
			&c_{\delta,\mG^{\theta}_{\sigma}}(a) &&= \int^a \frac{da'}{a'}c_\delta(a,a')\frac{D_{+}(a')}{D_{+}(a)}\int_{a'}^a da''D'_{+}(a'')&&\frac{D_{+}(a'')}{D_{+}(a)^2}\mG^{\theta}_{\sigma}(a'')\\ \nonumber
			&{}&&=\left(\mV^{\delta}_{\sigma1}(a)+\mV^{\delta}_{\sigma2}(a)\right)c_{\delta,1}-c_{\delta,\mV_{\sigma1}^\delta}-c_{\delta,\mV_{\sigma 2}^\delta}\\ \nonumber
			&c_{\psi\mG^{\delta}_{\sigma}}(a) &&= \int^a \frac{da'}{a'}c_\psi(a,a')\frac{D_{+}(a')^3}{D_{+}(a)^3}\mG^{\delta}_{\sigma}(a')\big(\mG^{\delta}_{1}(a')&&-\mG^{\theta}_{1}(a')\big)\\ \nonumber
			&c_{\delta,{123}}(a) &&= \int^a \frac{da'}{a'}c_{\delta}(a,a')\bigg[ \frac{1}{2}\frac{D_{+}(a')}{D_{+}(a)}- \frac{D_{+}(a')^2}{D_{+}(a)^2}&&+\frac{1}{2}\frac{D_{+}(a')^3}{D_{+}(a)^3}\bigg].
			\end{alignat}
			The integral that appears in $c_{\delta,\mG^{\theta}_{\sigma}}$ was solved using (\ref{tintegral}).
			\pagebreak

			As has been pointed out in section~\ref{stochred} the coefficients in (\ref{third order final coefficients}) have degeneracies. We here derive one of the degeneracies explicitly
			\bea
			c_{\beta_1,(3)}+c_{\beta_2,(3)}+c_{\gamma_1,(3)}+c_{\gamma_2,(3)} &=& c_{\delta,\mG^{\theta}_{1}} + c_{\delta,\mG^{\theta}_{2}} +c_{\delta,\mV^{\delta}_{11}} + c_{\delta,\mV^{\delta}_{12}} +c_{\delta,\mV^{\delta}_{21}} + c_{\delta,\mV^{\delta}_{22}}\\ \nonumber 
			&&- \left(c_{\psi,\mV^{\delta}_{11}} + c_{\psi,\mV^{\delta}_{12}} +c_{\psi,\mV^{\delta}_{21}} + c_{\psi,\mV^{\delta}_{22}}\right)\\ \nonumber 
			&&+ \left(c_{\psi,\mV^{\theta}_{11}} + c_{\psi,\mV^{\theta}_{12}} +c_{\psi,\mV^{\theta}_{21}} + c_{\psi,\mV^{\theta}_{22}}\right)\\ \nonumber 
			&& =	\fr{1}{2}c_{\delta,1},
			\eea
			the other ones are derived similarly. The above holds because of the identities in (\ref{eq:timeidents}), which are derived in Appendix~\ref{appendixa}.
			
			Furthermore, using the coefficients in (\ref{third order final coefficients}) we can derive $Y(a)$ as given in (\ref{Ynice}). Using the definition (\ref{Ydef}), we get 
			\bea
			Y(a)c_{\delta,1} = -\fr{3}{14}c_{\delta,1}+c_{\delta,\mG_1^\theta}+c_{\delta,\mV_{11}^\delta}+c_{\delta,\mV_{12}^\delta}.
			\eea Simply plugging in $c_{\delta,\mG_1^\theta}$ as given in (\ref{coefficientraw}), then gives 
			\bea
			Y(a) = -\fr{3}{14}+\mV^{\delta}_{11}(a)+\mV^{\delta}_{12}(a).
			\eea
			Finally, after capturing all the degeneracies, we can define the new basis given by the $\mathbb{C}_i$ operators that appear in (\ref{eq:delta_h_CoI_t})
			\bea \label{eq:Ci_ops}
				{}^*\mathbb{C}^{(1)}_{\delta} (\vq_1)&=& 1\\ \nonumber
				{}^*\mathbb{C}^{(2)}_{\delta} (\vq_1,\vq_2)&=& \beta(\vq_1,\vq_2)\\ \nonumber
				{}^*\mathbb{C}^{(2)}_{\alpha}(\vq_1,\vq_2)&=& \alpha(\vq_1,\vq_2)-\beta(\vq_1,\vq_2)\\ \nonumber 
				{}^*\mathbb{C}^{(2)}_{\mathbb{I}} (\vq_1,\vq_2)&=& 1\\ \nonumber
				{}^*\mathbb{C}^{(3)}_{\delta}(\vq_1,\vq_2,\vq_3)&=& -\fr{3}{14}\alpha_1(\vq_1,\vq_2,\vq_3)+\fr{3}{7}\alpha_2(\vq_1,\vq_2,\vq_3) +\fr{2}{7}\beta_2(\vq_1,\vq_2,\vq_3) + \fr{3}{14}\gamma_1(\vq_1,\vq_2,\vq_3)\\ \nonumber
				{}^*\mathbb{C}^{(3)}_{\alpha_1}(\vq_1,\vq_2,\vq_3)&=&\alpha_1(\vq_1,\vq_2,\vq_3)-\alpha_2(\vq_1,\vq_2,\vq_3)\\ \nonumber
				{}^*\mathbb{C}^{(3)}_{\beta_1}(\vq_1,\vq_2,\vq_3)&=&-\alpha_2(\vq_1,\vq_2,\vq_3) + \beta_1(\vq_1,\vq_2,\vq_3) - \gamma_1(\vq_1,\vq_2,\vq_3)\\ \nonumber
				{}^*\mathbb{C}^{(3)}_{\gamma_2}(\vq_1,\vq_2,\vq_3)&=&-\alpha_1(\vq_1,\vq_2,\vq_3)+2\alpha_2(\vq_1,\vq_2,\vq_3) - \beta_2(\vq_1,\vq_2,\vq_3) + \gamma_2(\vq_1,\vq_2,\vq_3)\\ \nonumber
				{}^*\mathbb{C}^{(3)}_{\alpha}(\vq_1,\vq_2,\vq_3)&=&\alpha(\vq_1,\vq_2)-\beta(\vq_1,\vq_2)\\ \nonumber
				{}^*\mathbb{C}^{(3)}_{\beta}(\vq_1,\vq_2,\vq_3)&=&\beta(\vq_1,\vq_2)\\ \nonumber
				{}^*\mathbb{C}^{(3)}_{\mathbb{I}}(\vq_1,\vq_2,\vq_3)&=&1\\ \nonumber
				{}^*\mathbb{C}^{(3)}_{Y}(\vq_1,\vq_2,\vq_3)&=&-\alpha_1(\vq_1,\vq_2,\vq_3)+2\alpha_2(\vq_1,\vq_2,\vq_3) - \beta_2(\vq_1,\vq_2,\vq_3) + \gamma_1(\vq_1,\vq_2,\vq_3),\\ \nonumber
				\eea
			where we used
			\be
			X^{(n)}(\vk,a) = \int\frac{d^3q_1}{(2\pi)^{3}}\ldots\frac{d^3q_n}{(2\pi)^{3}}(2\pi)^{3}\delta_{D}(\vk-\vq_1-\ldots-\vq_n) \ {}^*X^{(n)}(\vq_1,...,\vq_n)\ \delta^{(1)}_{\vq_1}(a)\ldots\delta^{(1)}_{\vq_n}(a).
			\ee
			\section{Derivig flow terms}\label{appendixc}
			We here give a few examples of how to derive the flow terms in (\ref{eq:euler_bias_new}) coming from the Taylor expansion (\ref{eq:xfl_expansion}). The Taylor expansion of $\delta(\xfl,a)$ is given by
			\bea\label{fluidtaylor}
			&&\delta(\xfl(a,a'),a')=\delta(\vec x,a')-\d_i\delta(x,a')\int_{a'}^a 	\fr{da''}{a''^2H(a'')} \; v^i(\vec x,a'')\\ \nonumber
			&&\quad \quad\quad\quad\quad\quad+\frac{1}{2}\d_i\d_j \delta(x,a')\int_{a'}^a 	\fr{da''}{a''^2H(a'')}\; v^i(\vec x,a'')\int_{a'}^a 	\fr{da'''}{a'''^2H(a''')}\;v^j(\vec x,a''')\\ \nonumber
			&&\quad \quad\quad \quad\quad \quad+\d_i\delta(x,a')\int_{a'}^a 	\fr{da''}{a''^2H(a'')}\; \d_j v^i(\vec x,a'') \int^a_{a''} 	\fr{da'''}{a'''^2H(a''')}\;v^j(\vec x,a''')+\ldots\ .
			\eea 
			After expanding the overdensity and velocity divergence perturbatively, the only second order term (apart from $\delta^{(2)}$) is in the first line, which is given by
			\bea
			&& -\int^a \fr{da'}{a'}\; c_{\delta}(a,a')\; \d_i \delta^{(1)}(a') \int^a_{a'} \fr{da''}{a''^2H(a'')} v^{(1)}{}^i(a'')=\\ \nonumber
			&&\quad=\int^a \fr{da'}{a'}\; c_{\delta}(a,a')\; \frac{D_{+}(a')}{D_{+}(a)}\d_i \delta^{(1)}(a) \int^a_{a'} da''\;\frac{D'_{+}(a'')}{D_{+}(a)}\frac{\d^i}{\d^2}\theta^{(1)} (a)\\ \nonumber
			&&\quad=\int^a \fr{da'}{a'}\; c_{\delta}(a,a')\; \frac{D_{+}(a')}{D_{+}(a)}\left[1-\frac{D_{+}(a')}{D_{+}(a)}\right]\d_i \delta^{(1)}(a) \frac{\d^i}{\d^2}\theta^{(1)} (a)\\ \nonumber
			&&\quad=\left[c_{\delta,1}(a)-c_{\delta,2}(a)\right]\d_i \delta^{(1)}(a) \frac{\d^i}{\d^2}\theta^{(1)} (a)\ .\\ \nonumber
			&&\quad=c_{\delta,12}(a)\d_i \delta^{(1)}(a) \frac{\d^i}{\d^2}\theta^{(1)} (a)\ .
			\eea
			At third order we take this same term with $\delta$ at second order and in $v$ at first order. Trivially, this gives
			\bea
			&& -\int^a \fr{da'}{a'}\; c_{\delta}(a,a')\; \d_i \delta^{(2)}(a') \int^a_{a'} \fr{da''}{a''^2H(a'')}\; v^{(1)}{}^i(a'')=\\ \nonumber
			&&\quad=\int^a \fr{da'}{a'}\; c_{\delta}(a,a')\; \left[1-\frac{D_{+}(a')}{D_{+}(a)}\right]\d_i \delta^{(2)}(a') \frac{\d^i}{\d^2}\theta^{(1)} (a)\\ \nonumber
			&&\quad\stackrel{\text{EdS}}{=}\left[c_{\delta,2}(a)-c_{\delta,3}(a)\right]\d_i \delta^{(2)}(a) \frac{\d^i}{\d^2}\theta^{(1)} (a)
			\eea
			Again, from the same term we can take $\delta$ at linear and $v$ at second order. We have
			\bea
			&& -\int^a \fr{da'}{a'}\; c_{\delta}(a,a')\; \d_i \delta^{(1)}(a') \int^a_{a'}\fr{da''}{a''^2H(a'')}\; v^{(2)}{}^i(a'')=\\ \nonumber
			&&\quad=\int^a \fr{da'}{a'}\; c_{\delta}(a,a')\; \frac{D_{+}(a')}{D_{+}(a)} \int^a_{a'} da''\;\frac{D'_{+}(a'')}{D_{+}(a'')}\d_i \delta^{(1)}(a)\frac{\d^i}{\d^2}\theta^{(2)} (a'')\\ \nonumber
			&&\quad\stackrel{\text{EdS}}{=}\fr{1}{2}\left[c_{\delta,1}(a)-c_{\delta,3}(a)\right]\d_i \delta^{(1)}(a) \frac{\d^i}{\d^2}\theta^{(2)} (a),
			\eea
			where we partially integrated to obtain the EdS results.
			In the second and third lines of (\ref{fluidtaylor}) we can take all fields at linear order. The two terms are derived in a similar fashion. For the third line for example, we have
			\bea
			&&- \frac{1}{2}\int^a \fr{da'}{a'}\; c_{\delta}(a,a')\; \d_i \delta^{(1)}(a') \int^a_{a'} \fr{da''}{a''^2H(a'')}\; \d_jv^{(1)}{}^i(a'') \int^a_{a''} \fr{da'''}{a'''^2H(a''')}\; v^{(1)}{}^j(a''')\\ \nonumber
			&&\quad=- \frac{1}{2}\int^a \fr{da'}{a'}\; c_{\delta}(a,a')\; \frac{D_{+}(a')}{D_{+}(a)}\d_i \delta^{(1)}(a) \int^a_{a'} da''\;\frac{D'_{+}(a'')}{D_{+}(a)} \int^a_{a''} da'''\;\frac{D'_{+}(a''')}{D_{+}(a)}\frac{\d_j\d^i}{\d^2}\theta^{(1)} (a)\frac{\d^j}{\d^2}\theta^{(1)} (a)\\ \nonumber
			&&\quad=- \frac{1}{2}\int^a \fr{da'}{a'}\; c_{\delta}(a,a')\; \frac{D_{+}(a')}{D_{+}(a)}\left[\frac{1}{2}-\frac{D_{+}(a')}{D_{+}(a)}+\frac{1}{2}\frac{D_{+}(a')^2}{D_{+}(a)^2}\right]\d_i \delta^{(1)}(a) \frac{\d^j\d^i}{\d^2}\theta^{(1)} (a) \frac{\d^j}{\d^2}\theta^{(1)}(a) \\ \nonumber
			&&\quad=-\frac{1}{2}c_{\delta,123}\d_i \delta^{(1)}(a) \frac{\d^i\d^j}{\d^2}\theta^{(1)} (a)\frac{\d^j}{\d^2}\theta^{(1)} (a)\, 
			\eea
			and the rest of the flow terms in (\ref{eq:euler_bias_new}) are derived similarly.

			\section{Relation to BoD basis}\label{appendixd}
			For completeness we here give the relation to the BoD operators $\tilde{\mathbb{C}}_i$ from~\cite{Perko:2016puo} (in~\cite{Perko:2016puo} $\tilde{\mathbb{C}}_i$ are called ${\mathbb{C}}_i$). They are related to the ones we used in this paper by
			
					\bea \label{eq:old_paper}
			\tilde{\mathbb{C}}^{(1)}_{\delta,1}(\vec k,a)&=&\mathbb{C}^{(1)}_{\delta}(\vec k,a)\\ \nonumber
			\tilde{\mathbb{C}}^{(2)}_{\delta,1}(\vec k,a)&=&\mathbb{C}^{(2)}_{\alpha}(\vec k,a)+\mathbb{C}^{(2)}_{\delta}(\vec k,a)-\mathbb{C}^{(2)}_{\mathbb{I}}(\vec k,a)\\ \nonumber
			\tilde{\mathbb{C}}^{(3)}_{\delta,1}(\vec k,a)&=&\mathbb{C}^{(3)}_{\alpha_1}(\vec k,a)+\fr{2}{7}\mathbb{C}^{(3)}_{\gamma_2}(\vec k,a)+\mathbb{C}^{(3)}_{\delta}(\vec k,a)-\fr{12}{7}\mathbb{C}^{(3)}_{\alpha}(\vec k,a)-2\mathbb{C}^{(3)}_{\beta}(\vec k,a)+\mathbb{C}^{(3)}_{\mathbb{I}}(\vec k,a)\\ \nonumber
			\tilde{\mathbb{C}}^{(2)}_{\delta,2}(\vec k,a)&=&-\fr{2}{7}\mathbb{C}^{(2)}_{\alpha}(\vec k,a)+\mathbb{C}^{(2)}_{\mathbb{I}}(\vec k,a)\\ \nonumber
			\tilde{\mathbb{C}}^{(3)}_{\delta,2}(\vec k,a)&=& -\fr{2}{7}\mathbb{C}^{(3)}_{\alpha_1}(\vec k,a)+\fr{16}{7}\mathbb{C}^{(3)}_{\alpha}(\vec k,a)+2\mathbb{C}^{(3)}_{\beta}(\vec k,a)-2\mathbb{C}^{(3)}_{\mathbb{I}}(\vec k,a)\\ \nonumber
			\tilde{\mathbb{C}}^{(3)}_{\delta,3}(\vec k,a)&=& \fr{1}{21} \mathbb{C}^{(3)}_{\beta_1}(\vec k,a) - \fr{4}{63}\mathbb{C}^{(3)}_{\gamma_2}(\vec k,a)-\fr{4}{7}\mathbb{C}^{(3)}_{\alpha}(\vec k,a)+\mathbb{C}^{(3)}_{\mathbb{I}}(\vec k,a)\\ \nonumber
			\tilde{\mathbb{C}}^{(2)}_{\delta^2,1}(\vec k,a)&=&\mathbb{C}^{(2)}_{\mathbb{I}}(\vec k,a)\\ \nonumber
			\tilde{\mathbb{C}}^{(3)}_{\delta^2,1}(\vec k,a)&=&2\mathbb{C}^{(3)}_{\alpha}(\vec k,a)+2\mathbb{C}^{(3)}_{\beta}(\vec k,a)-2\mathbb{C}^{(3)}_{\mathbb{I}}(\vec k,a)\\ \nonumber
			\tilde{\mathbb{C}}^{(3)}_{\delta^2,2}(\vec k,a)&=&-\fr{4}{7}\mathbb{C}^{(3)}_{\alpha}(\vec k,a)+2\mathbb{C}^{(3)}_{\mathbb{I}}(\vec k,a)\\ \nonumber
			\tilde{\mathbb{C}}^{(3)}_{s^2,2}(\vec k,a)&=& \fr{5}{7} \mathbb{C}^{(3)}_{\beta_1}(\vec k,a) - \fr{2}{7}\mathbb{C}^{(3)}_{\gamma_2}(\vec k,a)-\fr{29}{21}\mathbb{C}^{(3)}_{\alpha}(\vec k,a)+\fr{4}{3}\mathbb{C}^{(3)}_{\mathbb{I}}(\vec k,a)\\ \nonumber
			\tilde{\mathbb{C}}^{(3)}_{\delta^3,1}(\vec k,a)&=&\mathbb{C}^{(3)}_{\mathbb{I}}(\vec k,a).\\ \nonumber
			\eea
			Similarly, the bias coefficients in this paper $\{b_1,b_2,b_3,b_4\}$ are related to the ones in~\cite{Perko:2016puo} $\{\tilde{b}_1,\tilde{b}_2,\tilde{b}_3,\tilde{b}_4\}$ by
			\begin{eqnarray}
			\label{coefftransform}
			\tilde{b}_1 &=& b_1\\ \nonumber
			\tilde{b}_2 &=& \sfrac{7}{2}b_1-\sfrac{7}{2}b_3 \\ \nonumber
			\tilde{b}_3 &=& 21 b_4 \\ \nonumber
			\tilde{b}_4 &=& -\sfrac{5}{2}b_1+b_2+\sfrac{7}{2}b_3 
			\end{eqnarray}

			\section{Redshift space kernels}\label{appendixe}
			We here give explicit expressions for the integrals in equation (\ref{brackets}). 
			
			At second order we have
			\begingroup
			\allowdisplaybreaks
			\bea			\label{rsec}
			^*[\frac{\partial_z}{\partial^2} \theta_h \delh]^{(2)} (a)
			&=& \left( \frac{-i q_{1z}} {q_1^2} \right) K_{\theta_h} ^{(1)}(a)K_{\delta_h} ^{(1)}(a) \\ \nonumber
			^*[\frac{\partial_z}{\partial^2} \theta_h \frac{\partial_z}{\partial^2} \theta_h]^{(2)} (a)
			&=& \left( -\frac{q_{1z}q_{2z}}{q_1^2 q_2^2} \right) K_{\theta_h} ^{(1)}(a) K_{\theta_h} ^{(1)}(a) \ ,\\ \nonumber
			\\ \nonumber
			\\ \nonumber
			\text{and at third order}\\ \nonumber
			^*[\frac{\partial_z}{\partial^2} \theta_h \frac{\partial_z}{\partial^2} \theta_h \frac{\partial_z}{\partial^2} \theta_h]^{(3)} (a)	&=& \left(i \frac{ q_{1z}q_{2z}q_{3z}}{q_1^2 q_2^2 q_3^2} \right)K_{\theta_h} ^{(1)} (a)K_{\theta_h} ^{(1)}(a)K_{\theta_h} ^{(1)}(a) \\ \nonumber
			^*[\frac{\partial_z}{\partial^2} \theta_h \frac{\partial_z}{\partial^2} \theta_h \delh]^{(3)} (a) &=& \left(-\frac{ q_{1z}q_{2z}}{q_1^2 q_2^2} \right)K_{\theta_h} ^{(1)}(a) K_{\theta_h} ^{(1)}(a) K_{\delta_h} ^{(1)}(a) \\ \nonumber
			^*[\frac{\partial_z}{\partial^2} \theta_h \delh]^{(3)} (a) &=& \ \left( -\frac{i q_{3z}}{q_3^2}K_{\delta_h} ^{(2)}(\q_1, \q_2,a) K_{\theta_h} ^{(1)}(a)- \frac{i (\q_1+\q_2)_z}{(\q_1+\q_2)^2}K_{\theta_h} ^{(2)}(\q_1, \q_2,a) K_{\delta_h} ^{(1)}(a) \right) \\ \nonumber
			^*[\frac{\partial_z}{\partial^2} \theta_h \frac{\partial_z}{\partial^2} \theta_h]^{(3)} (a) 	&=& \left( -\frac{(\q_1+\q_2)_z q_{3z}}{(\q_1+\q_2)^2 \q_3^2} \right)2K_{\theta_h} ^{(2)}(\q_1,\q_2,a)K_{\theta_h}^{(1)}(a),
			\eea
			\endgroup
			where we again used the notation
			\bea
			[X]^{(n)}_{\vk}(a) &=& \int\frac{d^3q_1}{(2\pi)^{3}}\ldots\frac{d^3q_n}{(2\pi)^{3}}(2\pi)^{3}\delta_{D}(\vk-\vq_1-\ldots-\vq_n) \ {}^*[X]^{(n)}(a) \delta^{(1)}_{\vq_1}(a)\ldots\delta^{(1)}_{\vq_n}(a).
			\eea

			\section{Further EdS comparisons}\label{appendixf}
\begin{figure}[h!]
				\begin{center}
					\includegraphics[width=17cm]{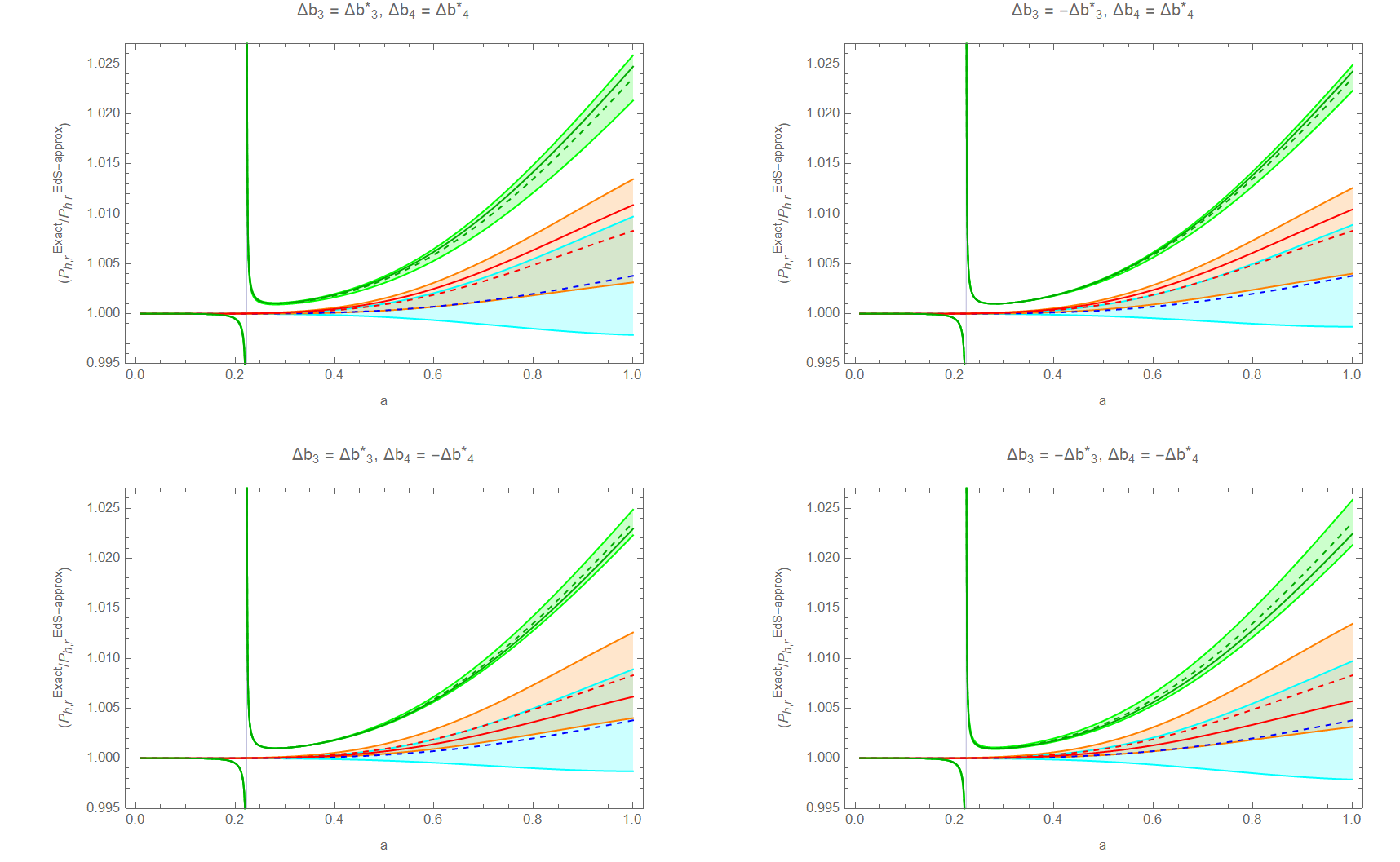}
					\caption{\label{fig:relatives}\footnotesize Diagrammatic representation of the ratio of the exact galaxy power spectrum in redshift space over the approximate case as a function of the scale factor at $k^* = 0.2\ \text{hMpc}^{-1}$. The plots show the ratios of the real parts $P_{\rm Real}^{\text{Exact}}(k^*, a)/P_{\rm Real}^{\text{EdS-approx}}(k^*, a)$ (blue/cyan), the monopoles $P_{0}^{\text{Exact}}(k^*,a)/P_{0}^{\text{EdS-approx}}(k^*,a)$ (red/orange) and the quadrupoles $P_{2}^{\text{Exact}}(k^*,a)/P_{2}^{\text{EdS-approx}}(k^*,a)$ (dark/light green) of the galaxy power spectrum in redshift space. For the bias coefficients with EdS approximation we used $b_{1,\text{EdS}}(a^*) = 2.2$, $b_{2,\text{EdS}}(a^*) = -0.4$, $b_{3,\text{EdS}}(a^*) = 1.9$, $b_{4,\text{EdS}}(a^*) = 0$, $c_{ct,\text{EdS}}(a^*) = 0$, $\tilde{c}_{r,1,\text{EdS}}(a^*) = -8\left(\text{k}_\text{M}/\text{hMpc}^{-1}\right)^2$, $\tilde{c}_{r,2,\text{EdS}}(a^*) = 0,\quad c_{\epsilon,1,\text{EdS}}(a^*) = 1.3$ and $c_{\epsilon,2,\text{EdS}}(a^*) = -4\left(\text{k}_\text{M}/\text{hMpc}^{-1}\right)^2$ from~\cite{DAmico:2019fhj} at $a^* = 0.64$. The coefficients were promoted to functions through the time dependence implied by (\ref{eq:greenestimate}). Furthermore, we use the calculable time dependence of $Y(a)$ from (\ref{Ynice}). The dashed lines represent the effect of the approximation that comes from redshift space and the contribution multiplied by $Y(a)$ only, i.e. $b_3(a) = b_{3,\text{EdS}}(a)$ and $ b_4(a) = b_{4,\text{EdS}}(a)$. The estimate from (\ref{eq:greenestimate}), where $b_3(a) = b_{3,\text{EdS}}(a)+\Delta b_3^*(a)$ and $ b_4(a) = b_{4,\text{EdS}}(a)+\Delta b_4^*(a)$ ($\Delta b_3^*(a)$ and $\Delta b_4^*(a)$ are shown in Figure~\ref{fig:biasdif}), is depicted by the darker solid lines. The lighter shaded areas are bounded from below ($-$) and above ($+$) by $b_3(a)= b_{3,\text{EdS}}(a)\pm2*\Delta b_3^*(a)$ and $b_4(a)= b_{4,\text{EdS}}(a)\pm2*\Delta b_4^*(a)$. In each diagram we used $(\Delta b_3(a),\Delta b_4(a))= (\pm\Delta b_3^*(a),\pm\Delta b_4^*(a))$, where the specific configuration is given in the title of each figure and $\Delta b_3^*(a^*)$ and $\Delta b_4^*(a^*)$ are the estimates from (\ref{eq:greenestimate}).}
				\end{center}
			\end{figure}

			In the main text we used $c_g = 0.58$ to calculate (\ref{eq:greenestimate}), and apply it for $\Delta b_3^*$ and $\Delta b_4^*$. It is interesting to see how a different sign of these estimates would affect our result.
			Therefore, in the first part of this Appendix, we here give plots of Figure~\ref{fig:evolution} with all possible combinations of the relative signs of $\Delta b_3$ and $\Delta b_4$ (four possible combinations in total). They are depicted in Figure~\ref{fig:relatives}. \pagebreak
		
			\pagebreak
			
\begin{figure}[h!]
				\begin{center}
					\begin{minipage}[b]{0.45\textwidth}
						\includegraphics[width=8 cm,height = 19cm]{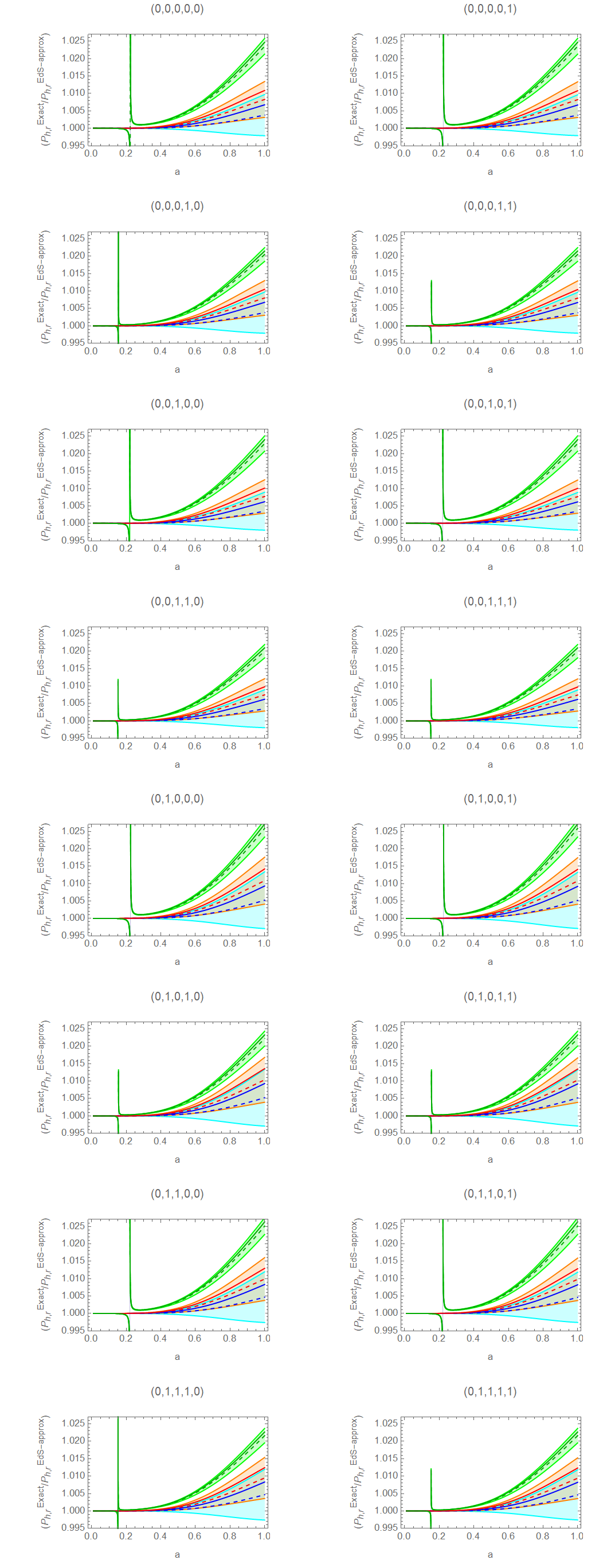}
						
					\end{minipage}
					\begin{minipage}[b]{0.45\textwidth}
						\includegraphics[width=8 cm,height = 19cm]{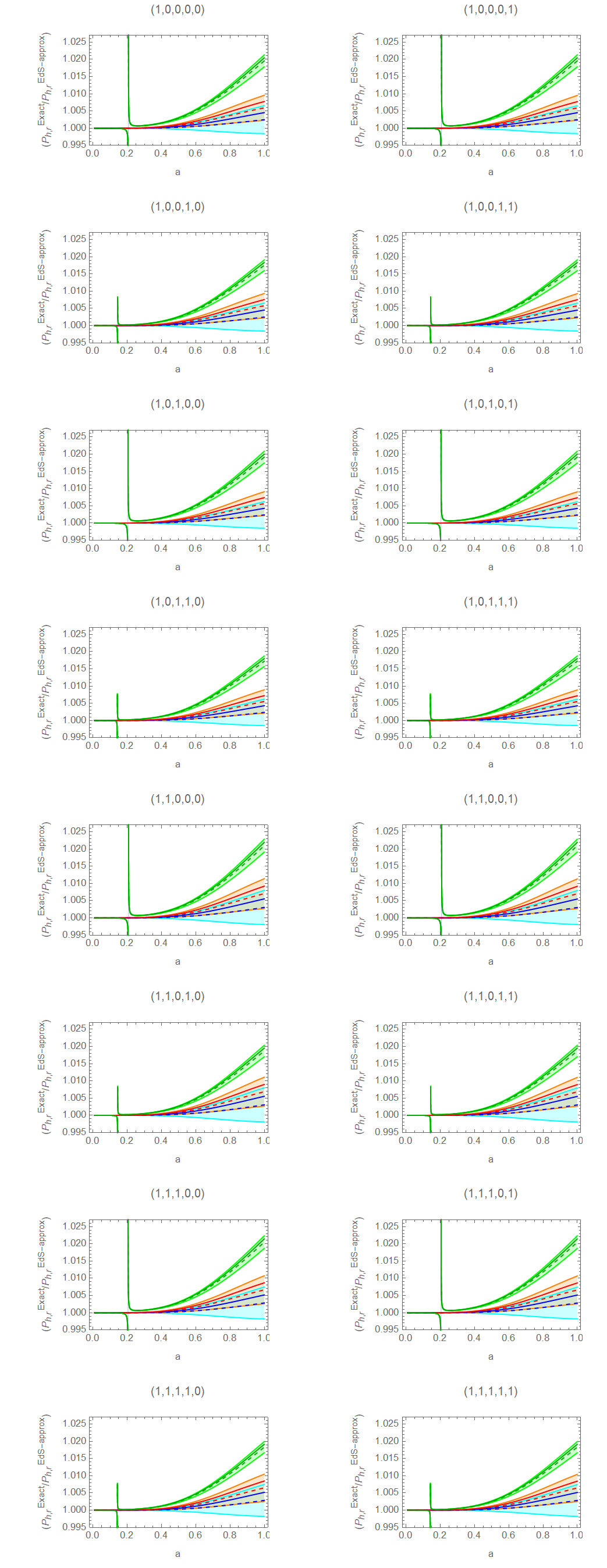}
						
					\end{minipage}
					\caption{\label{fig:allpos1}\footnotesize Plotted above are the ratios of the exact galaxy power spectrum in redshift space over the approximate case as a function of the scale factor at $k^* = 0.2\ \text{hMpc}^{-1}$. The plots show the ratios of the real parts $P_{\rm Real}^{\text{Exact}}(k^*, a)/P_{\rm Real}^{\text{EdS-approx}}(k^*, a)$ (blue/cyan), the monopoles $P_{0}^{\text{Exact}}(k^*,a)/P_{0}^{\text{EdS-approx}}(k^*,a)$ (red/orange) and the quadrupoles $P_{2}^{\text{Exact}}(k^*,a)/P_{2}^{\text{EdS-approx}}(k^*,a)$ (dark/light green) of the galaxy power spectrum in redshift space. For the bias coefficients with EdS approximation we used $b_{1,\text{EdS}}(a^*) = 2.2$, $b_{2,\text{EdS}}(a^*) = -0.4$, $b_{3,\text{EdS}}(a^*) = 1.9$, $b_{4,\text{EdS}}(a^*) = 0$, $c_{ct,\text{EdS}}(a^*) = 0$, $\tilde{c}_{r,1,\text{EdS}}(a^*) = -8\left(\text{k}_\text{M}/\text{hMpc}^{-1}\right)^2$, $\tilde{c}_{r,2,\text{EdS}}(a^*) = 0,\quad c_{\epsilon,1,\text{EdS}}(a^*) = 1.3$ and $c_{\epsilon,2,\text{EdS}}(a^*) = -4\left(\text{k}_\text{M}/\text{hMpc}^{-1}\right)^2$ from~\cite{DAmico:2019fhj} at $a^* = 0.64$. Here $(db_2,db_3,dc_{ct},d\tilde{c}_{r,1},dc_{\epsilon,1})\in \{0,1\}^5$ and the particular choice is in the title of each figure. The further procedure and color code is the same as in Figure~\ref{fig:evolution}.}
					
				\end{center}
			\end{figure}

			Next, we want to check that our results do not depend too much on the specific choice of bias coefficients we used. This is important since
			the coefficients measured in~\cite{DAmico:2019fhj} have quite large error bars. Approximately we have
			\bea
			&&b_{1,\text{EdS}} = 2.2\pm0.2 \quad 	b_{2,\text{EdS}} = -0.4\pm0.5 \quad 	b_{3,\text{EdS}} = 1.9\pm0.2 \quad 	b_{4,\text{EdS}} = 0\pm0.14 \quad \\ \nonumber 	
			&&c_{ct,\text{EdS}} = 0\pm3 \left(\fr{\text{k}_\text{NL}}{\text{hMpc}^{-1}}\right)^2\quad
			\tilde{c}_{r,1,\text{EdS}} = -8\pm4\left(\fr{\text{k}_\text{M}}{\text{hMpc}^{-1}}\right)^2 \quad 	\tilde{c}_{r,2,\text{EdS}} = 0 \\ \nonumber 	
			&& 
			c_{\epsilon,1,\text{EdS}} = 1.3\pm0.8 \quad c_{\epsilon,2,\text{EdS}} = -4\pm2\left(\fr{\text{k}_\text{M}}{\text{hMpc}^{-1}}\right)^2.
			\eea
			Note, however, that the errors on some of the parameters are highly correlated and we can treat them as one. We have $c_{\epsilon,1,\text{EdS}}=-\fr{1}{3}c_{\epsilon,2,\text{EdS}}$ and $\tilde b_{2,\text{EdS}}= \tilde b_{4,\text{EdS}}$ ($\tilde b$ are the bias coefficients in the basis of~\cite{DAmico:2019fhj} and the transformation is given in (\ref{coefftransform})), since their difference was put to zero in~\cite{DAmico:2019fhj}. Furthermore, since $b_1$ defines the proportionality constant in the time kernel function we leave $b_1$ out of our analysis. Therefore, there are five parameters we vary. The plots show versions of Figure~\ref{fig:evolution} with all possible applications of these errors. Next, we define the array $(db_2,db_4,dc_{ct},d\tilde{c}_{r,1},dc_{\epsilon,1})\in \{0,1,-1\}^5$. In the first case, a zero means we use the parameter itself and a one means we use the parameter plus its error. There are 32 combinations of these errors which are shown in Figure~\ref{fig:allpos1}. 
			
\begin{figure}[h!]
				\begin{center}
					\begin{minipage}[b]{0.45\textwidth}
						\includegraphics[width=8 cm,height = 19cm]{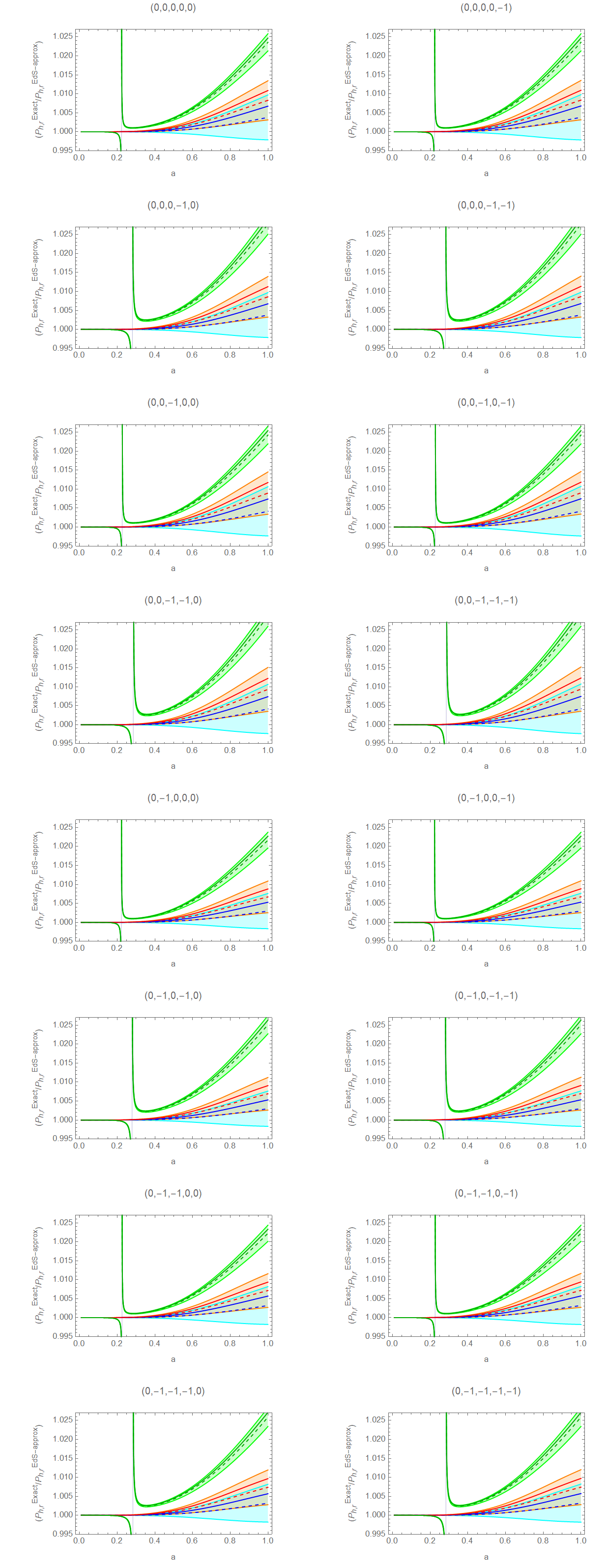}
						
					\end{minipage}
					\begin{minipage}[b]{0.45\textwidth}
						\includegraphics[width=8 cm,height = 19cm]{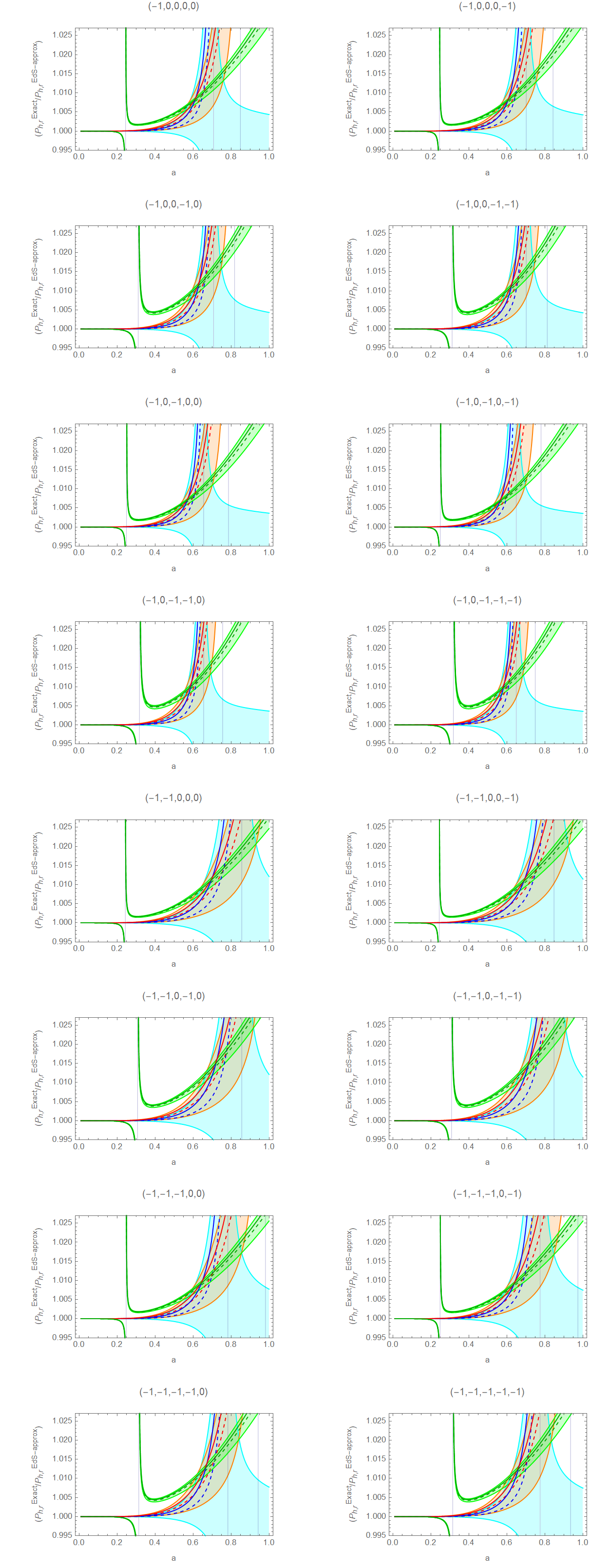}
						
					\end{minipage}
					\caption{\label{fig:allneg1}\footnotesize Plotted above are the ratios of the exact galaxy power spectrum in redshift space over the approximate case as a function of the scale factor at $k^* = 0.2\ \text{hMpc}^{-1}$. The plots show the ratios of the real parts $P_{\rm Real}^{\text{Exact}}(k^*, a)/P_{\rm Real}^{\text{EdS-approx}}(k^*, a)$ (blue/cyan), the monopoles $P_{0}^{\text{Exact}}(k^*,a)/P_{0}^{\text{EdS-approx}}(k^*,a)$ (red/orange) and the quadrupoles $P_{2}^{\text{Exact}}(k^*,a)/P_{2}^{\text{EdS-approx}}(k^*,a)$ (dark/light green) of the galaxy power spectrum in redshift space. For the bias coefficients with EdS approximation we used $b_{1,\text{EdS}}(a^*) = 2.2$, $b_{2,\text{EdS}}(a^*) = -0.4$, $b_{3,\text{EdS}}(a^*) = 1.9$, $b_{4,\text{EdS}}(a^*) = 0$, $c_{ct,\text{EdS}}(a^*) = 0$, $\tilde{c}_{r,1,\text{EdS}}(a^*) = -8\left(\text{k}_\text{M}/\text{hMpc}^{-1}\right)^2$, $\tilde{c}_{r,2,\text{EdS}}(a^*) = 0,\quad c_{\epsilon,1,\text{EdS}}(a^*) = 1.3$ and $c_{\epsilon,2,\text{EdS}}(a^*) = -4\left(\text{k}_\text{M}/\text{hMpc}^{-1}\right)^2$ from~\cite{DAmico:2019fhj} at $a^* = 0.64$. Here $(db_2,db_3,dc_{ct},d\tilde{c}_{r,1},dc_{\epsilon,1})\in \{0,-1\}^5$ and the particular choice is in the title of each figure. The further procedure and color code is as in Figure~\ref{fig:evolution}.}
					
				\end{center}
			\end{figure}

			Since we see no large difference in Figure~\ref{fig:allpos1}, which represents the possible addition of the error bar, we can treat the addition case as negligible. In the next plot, a zero means we use the parameter itself and a minus one means we use the parameter minus its error. There are 32 combinations of these errors shown in Figure~\ref{fig:allneg1}, where one can see that subtracting $db_2$ leads to an increase in the effect.

			\pagebreak

		\end{appendix}
		
		\FloatBarrier
		\bibliography{references}

	\end{document}